\documentclass[aps,prl,reprint,showpacs,groupedaddress,amsmath,amssymb,superscriptaddress,nofootinbib,amsmath]{revtex4-1}
\usepackage{graphicx}
\usepackage{dcolumn}
\usepackage{bm}
\usepackage{ulem}
\usepackage{amsmath}
\usepackage{xcolor}
\definecolor{apsblue}{rgb}{0.176, 0.152, 0.57}
\usepackage[colorlinks=true, linkcolor=black, citecolor=apsblue, filecolor=black, urlcolor=apsblue]{hyperref}
\usepackage{siunitx}
\usepackage{xfrac}
\begin{document}

\title{Spatiotemporal dynamics of ultrarelativistic beam-plasma instabilities}

\author{P.~San Miguel Claveria}
\email{pablo.san-miguel-claveria@polytechnique.edu}
\affiliation{LOA, ENSTA Paris, CNRS, Ecole Polytechnique, Institut Polytechnique de Paris, 91762 Palaiseau, France}
\author{X.~Davoine}
\affiliation{CEA, DAM, DIF, 91297 Arpajon, France}
\affiliation{Universit\'{e} Paris-Saclay, CEA, LMCE, 91680 Bruy\`{e}res-le-Ch\^{a}tel, France}
\author{J.~R.~Peterson}
\affiliation{SLAC National Accelerator Laboratory, Menlo Park, CA 94025, USA}
\affiliation{Stanford University, Physics Department, Stanford, CA 94305, USA}
\author{M.~Gilljohann}
\affiliation{LOA, ENSTA Paris, CNRS, Ecole Polytechnique, Institut Polytechnique de Paris, 91762 Palaiseau, France}
\author{I.~Andriyash}
\affiliation{LOA, ENSTA Paris, CNRS, Ecole Polytechnique, Institut Polytechnique de Paris, 91762 Palaiseau, France}
\author{R.~Ariniello}
\affiliation{University of Colorado Boulder, Department of Physics, Center for Integrated Plasma Studies, Boulder, Colorado 80309, USA}
\author{C.~Clarke}
\affiliation{SLAC National Accelerator Laboratory, Menlo Park, CA 94025, USA}
\author{H.~Ekerfelt}
\affiliation{SLAC National Accelerator Laboratory, Menlo Park, CA 94025, USA}
\author{C.~Emma}
\affiliation{SLAC National Accelerator Laboratory, Menlo Park, CA 94025, USA}
\author{J.~Faure}
\affiliation{CEA, DAM, DIF, 91297 Arpajon, France}
\affiliation{Universit\'{e} Paris-Saclay, CEA, LMCE, 91680 Bruy\`{e}res-le-Ch\^{a}tel, France}
\author{S.~Gessner}
\affiliation{SLAC National Accelerator Laboratory, Menlo Park, CA 94025, USA}
\author{M.~J.~Hogan}
\affiliation{SLAC National Accelerator Laboratory, Menlo Park, CA 94025, USA}
\author{C.~Joshi}
\affiliation{University of California Los Angeles, Los Angeles, CA 90095, USA}
\author{C.~H.~Keitel}
\affiliation{Max-Planck-Institut f\"ur Kernphysik, Saupfercheckweg 1, D-69117 Heidelberg, Germany}
\author{A.~Knetsch}
\affiliation{LOA, ENSTA Paris, CNRS, Ecole Polytechnique, Institut Polytechnique de Paris, 91762 Palaiseau, France}
\author{O.~Kononenko}
\affiliation{LOA, ENSTA Paris, CNRS, Ecole Polytechnique, Institut Polytechnique de Paris, 91762 Palaiseau, France}
\author{M.~Litos}
\affiliation{University of Colorado Boulder, Department of Physics, Center for Integrated Plasma Studies, Boulder, Colorado 80309, USA}
\author{Y.~Mankovska}
\affiliation{LOA, ENSTA Paris, CNRS, Ecole Polytechnique, Institut Polytechnique de Paris, 91762 Palaiseau, France}
\author{K.~Marsh}
\affiliation{University of California Los Angeles, Los Angeles, CA 90095, USA}
\author{A.~Matheron}
\affiliation{LOA, ENSTA Paris, CNRS, Ecole Polytechnique, Institut Polytechnique de Paris, 91762 Palaiseau, France}
\author{Z.~Nie}
\affiliation{University of California Los Angeles, Los Angeles, CA 90095, USA}
\author{B.~O'Shea}
\affiliation{SLAC National Accelerator Laboratory, Menlo Park, CA 94025, USA}
\author{D.~Storey}
\affiliation{SLAC National Accelerator Laboratory, Menlo Park, CA 94025, USA}
\author{N.~Vafaei-Najafabadi}
\affiliation{Stony Brook University, Stony Brook, NY 11794, USA}
\author{Y.~Wu}
\affiliation{University of California Los Angeles, Los Angeles, CA 90095, USA}
\author{X.~Xu}
\affiliation{SLAC National Accelerator Laboratory, Menlo Park, CA 94025, USA}
\author{J.~Yan}
\affiliation{Stony Brook University, Stony Brook, NY 11794, USA}
\author{C.~Zhang}
\affiliation{University of California Los Angeles, Los Angeles, CA 90095, USA}
\author{M.~Tamburini}
\affiliation{Max-Planck-Institut f\"ur Kernphysik, Saupfercheckweg 1, D-69117 Heidelberg, Germany}
\author{F.~Fiuza}
\affiliation{SLAC National Accelerator Laboratory, Menlo Park, CA 94025, USA}
\author{L.~Gremillet}
\email{laurent.gremillet@cea.fr}
\affiliation{CEA, DAM, DIF, 91297 Arpajon, France}
\affiliation{Universit\'{e} Paris-Saclay, CEA, LMCE, 91680 Bruy\`{e}res-le-Ch\^{a}tel, France}
\author{S.~Corde}
\email{sebastien.corde@polytechnique.edu}
\affiliation{LOA, ENSTA Paris, CNRS, Ecole Polytechnique, Institut Polytechnique de Paris, 91762 Palaiseau, France}

\date{\today}
\begin{abstract}

An electron or electron-positron beam streaming through a plasma is notoriously prone to micro-instabilities. For a dilute ultrarelativistic infinite beam, the dominant instability is a mixed mode between longitudinal two-stream and transverse filamentation modes, with a phase velocity oblique to the beam velocity. 
A spatiotemporal theory describing the linear growth of this oblique mixed instability
is proposed, which predicts that spatiotemporal effects generally prevail
for finite-length beams, leading to a significantly slower instability evolution than in the usually assumed purely temporal regime. These results are accurately supported by particle-in-cell (PIC) simulations.  
Furthermore,
we show that the self-focusing dynamics caused by the plasma wakefields driven by finite-width beams
can compete with the oblique instability.
Analyzed through PIC simulations, the interplay of these two processes in realistic systems bears important implications for upcoming accelerator experiments on ultrarelativistic beam-plasma interactions.
\end{abstract}
\maketitle
A large number of astrophysical and laboratory systems involve the collective interaction between beams of relativistic charged particles and plasmas. In many cases, this interaction is governed by plasma micro-instabilities which lead to electrostatic and electromagnetic fluctuations growing at kinetic scales, and mediating most of the energy and momentum transfers between the beam and plasma particles \cite{Sudan_Handbook_1984, Bret_POP_2010}. 

In astrophysics, these instabilities are thought to dissipate into heat or radiation the kinetic energy of relativistic outflows from various powerful sources (e.g. pulsar wind nebulae, neutron star mergers, active galactic nuclei). Notably, as a result of their nonlinear evolution~\cite{Bret_POP_2013}, they can spawn relativistic collisionless shock waves~\cite{Spitkovsky_APJ_2008a, *Lemoine_PRL_2019} which, in turn, are believed to generate the most energetic particles and radiations in the Universe~\cite{Blandford_PR_1987, *Bykov_AAR_2011}, including the electromagnetic counterpart of gravitational wave sources~\cite{Abbott_APJL_2017}. Beam-plasma instabilities therefore lie at the heart of the fast-emerging field of multi-messenger astrophysics~\cite{Meszaros_NRP_2019}. Another topic of active current research is their possibly crucial role in shaping the GeV photon emission from blazars, the microphysics of which remaining little understood \cite{Broderick_APJ_2012, *Sironi_APJ_2014, *Chang_APJ_2016}. 

Beyond their fundamental and astrophysical significance, these instabilities have a prominent place in experimental concepts utilizing relativistic beam-plasma interactions, such as staging of laser (LWFA) and plasma wakefield acceleration (PWFA)~\cite{Raj_PRR_2020}, or laser-driven ion acceleration~\cite{Fuchs_PRL_2003, Gode_PRL_2017}, against which they act detrimentally. Lately, it has also been proposed to harness them as a novel channel of $\gamma$-ray radiation~\cite{Benedetti_NP_2018}.
Now, progress in particle accelerators make it possible to envision probing these plasma processes in the laboratory~\cite{Allen_PRL_2012}. In particular, extreme beam parameters, with Lorentz factors $\gamma_b > 10^4$ and densities $n_b = 10^{18}-10^{20}\,\rm{cm}^{-3}$ will soon be available at the new Facility for Advanced Accelerator Tests II (FACET-II)~\cite{Yakimenko_PRAB_2019}. This will open unprecedented opportunities to investigate, under various plasma conditions and in a very controlled way, the effects of micro-instabilities on the beam propagation in the ultrarelativistic regime.

The micro-instabilities arising in a relativistic beam-plasma system
are usually classified into three types: the longitudinal two-stream instability (TSI), the transverse current filamentation instability (CFI) and the mixed mode, or oblique two-stream instability (OTSI) \cite{Bret_POP_2010, Lemoine_MNRAS_2010}. While several modes can develop simultaneously
from thermal noise or beam-induced perturbations,
a specific instability class generally dominates the early beam-plasma interaction.
A fully kinetic theory exists which describes the linear phase of the instability for unbounded (i.e., infinite) beam-plasma systems, allowing the dominant mode to be predicted for a given set of beam-plasma parameters~\cite{Bret_PRL_2008, Bret_POP_2010}.
A key finding is the dominance of the mixed mode over CFI and TSI in the case of ultrarelativistic ($\gamma_b \gg 1$) and dilute ($\alpha \equiv n_b/n_p \ll 1$, where $n_p$ is the electron plasma density) beams. This leads to density and field modulations with a longitudinal wavenumber $k_x \simeq  c/\omega_p \equiv k_p$ and a transverse wavenumber $k_\perp \gtrsim k_p$, growing at a maximum rate
\begin{equation}
    \Gamma_{\rm OTSI} = \frac{\sqrt{3}}{2^{4/3}} \left( \frac{1}{\gamma_b} \frac{n_b}{n_p} \frac{k_\perp^2}{k_p^2 + k_\perp^2} \right)^{1/3} \omega_p \,.
    \label{eq:G_OTSI}
\end{equation} 
where $\omega_p$ is the background plasma frequency, and $n_b$ is the sum of the number densities $n_{b\pm}$ of the beam electrons and positrons (if any). 
Still, this temporal theory cannot be directly applied to the finite-size beams or plasma boundaries involved in realistic settings, such as future high-energy accelerator experiments. The first attempts to account for inhomogeneity effects on linear beam-plasma instabilities concerned the TSI \cite{Bers_Handbook_1983, Jones_POF_1983}, revealing its pulse-shaped profile in case of localized initial disturbances. Recently, a model of the CFI excited by a longitudinally semi-infinite beam was proposed \cite{Pathak_NJP_2015}, showing that for moderate Lorentz factors ($\gamma_b \le 10$), spatiotemporal effects are present at the beam head. Interestingly, this model predicts spatiotemporal effects to vanish in the ultrarelativistic limit.

For oblique modes, thought to dominate for $\gamma_b \gg 1$ and $\alpha \ll 1$, no spatiotemporal theory exists~\cite{Shukla_NJP_2020}. Yet, from the above previous works and related studies of laser-plasma  \cite{Decker_PoP_1996, McKinstrie_PoP_1996} or beam-plasma \cite{Huang_PRL_2007,Kumar_PRL_2010} instabilities, one may expect finite beam dimensions --or, more generally, boundaries in the beam-plasma system-- to strongly impact the dynamics of the oblique modes.

In this Letter we address two phenomena arising when a relativistic beam of finite spatial extent is considered in a beam-plasma system subject to streaming instabilities.
First, we develop a spatiotemporal theory for the evolution of the OTSI, highlighting its spatiotemporal nature and resulting slower dynamics when a finite beam length is considered. 
Second, we show that the interplay of beam-plasma instabilities and the wakefield that is excited by a beam of finite length and width conveys constraints on the system parameters for the instabilities to dominate the interaction.
These results are particularly relevant to future accelerator experiments aiming to explore ultrarelativistic beam-plasma instabilities and their radiative by-products~\cite{Benedetti_NP_2018}.

\begin{figure}[t]
    \centering
    \includegraphics[width=0.5\textwidth]{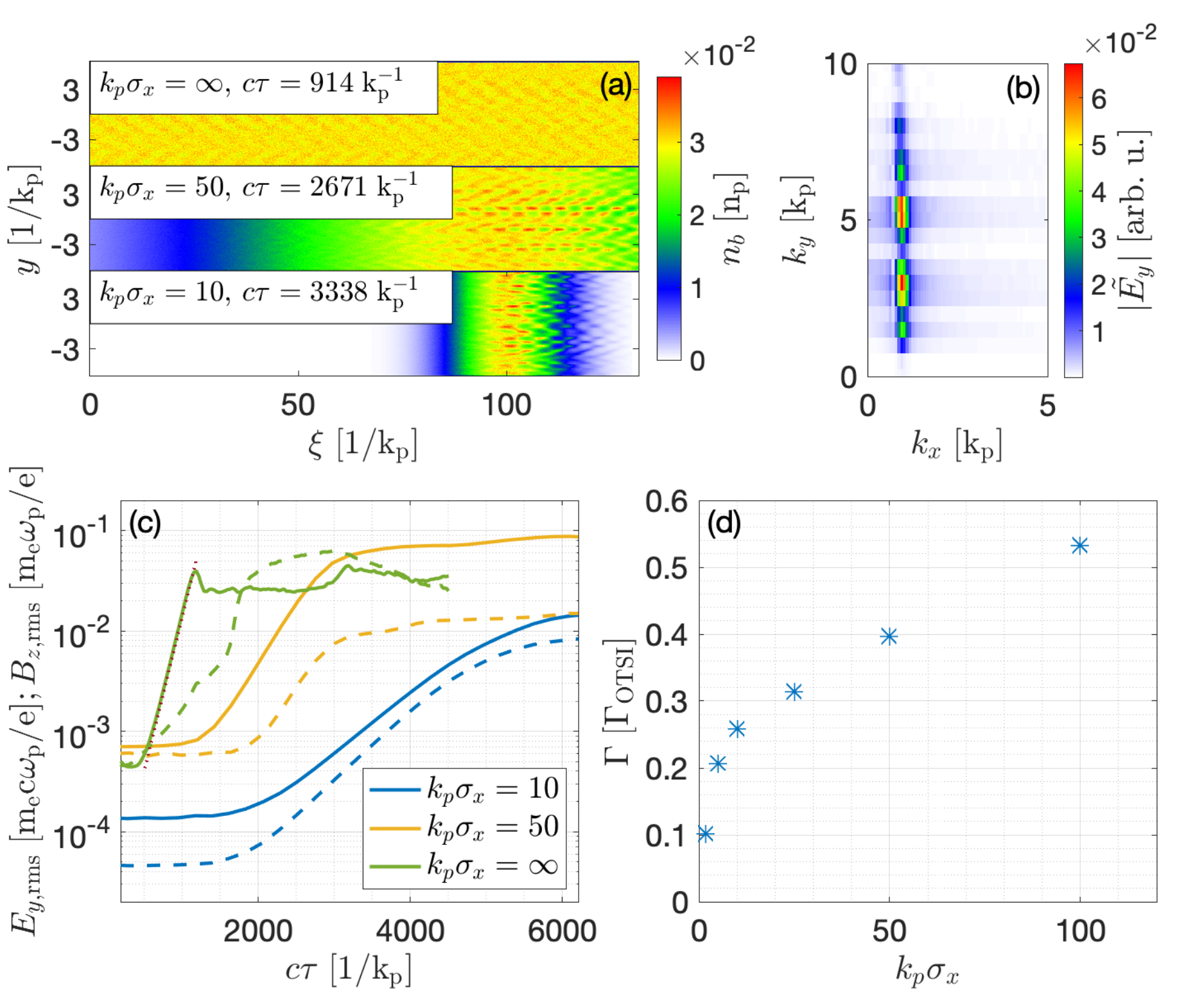}
    \caption{Simulated instability dynamics for ultrarelativistic ($\gamma_b=2\times 10^4$), dilute ($\alpha=0.03$) electron beams of various normalized lengths ($k_p \sigma_x$).
    (a) Snapshots of the beam density profile in the comoving coordinates $(\xi,\tau)=(v_bt-x,t)$ for different beam lengths.
    (b) 2D Fourier spectrum of the $E_y$ field fluctuations at $c\tau=3338 k_p^{-1}$ for $k_p \sigma_x =10$.  
    (c) Transverse electric field $E_{y,\mathrm{rms}}=\langle E_y^2\rangle^{1/2}$  (solid line) and magnetic field $B_{z,\mathrm{rms}}=\langle B_z^2\rangle^{1/2}$ (dashed line) averaged over $\xi\in [\xi_{\rm peak}-\sigma_x/2,\xi_{\rm peak}+\sigma_x/2]$ ($\xi_{\rm peak}$ the position of the beam center in the comoving coordinates) as a function of the beam propagation distance in the plasma $(c\tau)$ and the beam length. The dotted line plots the theoretical growth of the OTSI, Eq.~\eqref{eq:G_OTSI}, in the infinite beam case. The evaluation of the dominant $k_\perp$ in Eq.~\eqref{eq:G_OTSI} is carried out using the electrostatic result $\langle E_y^2\rangle/\langle E_x^2 \rangle \simeq (k_\perp/k_p)^2$.
    (d) Effective growth rate ($\Gamma/\Gamma_{\rm OTSI}$) vs. $k_p \sigma_x$ within the central slice of the beam (see text for details). 
    }
    \label{fig:fig1}
\end{figure}
We start by presenting the results of 2D PIC 
\textsc{calder} \cite{Lefebvre_NF_2003} simulations of an ultrarelativistic ($\gamma_b = 2\times10^4$), low-density ($\alpha=0.03$) electron beam interacting with a uniform electron-proton plasma.
The mesh size was set to $(\Delta x,\Delta y)= (0.042,0.084) k_p^{-1}$, the time step was $\Delta t = 0.041\omega_p^{-1}$, and 100 macroparticles per cell were used for each species (beam electrons, plasma electrons and ions). The beam profile was taken to be Gaussian in the longitudinal ($x$) direction with a RMS length of $\sigma_x$, and uniform in the transverse ($y$) direction. Unless otherwise mentioned, the boundary conditions were absorbing along $x$ and periodic along $y$, for both the fields and particles.

\begin{figure*}[ht!]
    \centering
    \includegraphics[width=\textwidth]{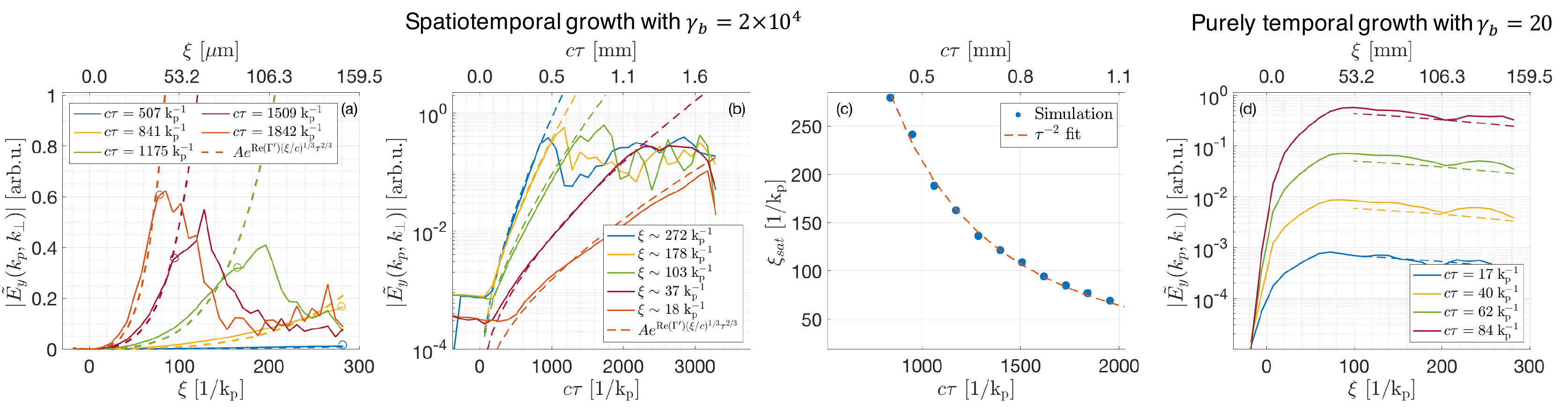}
    \caption{2D PIC simulations of the OTSI induced by a step-like $e^-e^+$ pair beam and comparison with linear theory in the spatiotemporal regime for $\gamma_b = 2\times 10^4$ (a-c) and the temporal regime for $\gamma_b = 20$ (d). (a) and (d): Spectral amplitude $\vert \widetilde{E}_y(k_x,k_\perp) \vert$ of the dominant oblique mode ($k_x=k_p$, $k_\perp\simeq 3k_p$) as a function of $\xi$ for different propagation distances $c\tau$. (b): Same quantity but as a function of $c\tau$ for different beam slices $\xi$. In (a) and (b), the simulation data (solid lines)  is fitted to the theoretical law $A\exp[(3/2^{2/3})\Gamma_{\rm OTSI} (\xi/c)^{1/3} \tau^{2/3}]$ for $\xi \le \xi_{\rm sat}$ (dashed lines). (c) Saturation position $\xi_{\rm sat}$ [also shown in (a) as circles] vs. $c\tau$ (filled circles), compared with the theoretical expectation $\xi_{\rm sat} \propto \tau^{-2}$ (red dashed line). Dashed lines in (d) plot the theoretical temporal growth of $\vert \widetilde{E}_y (k_p,3k_p)\vert$ at different times $c\tau \ge 17 \rm k_p^{-1}$
    .}
    \label{fig:fig2}
\end{figure*}

Figure~\ref{fig:fig1}(a) illustrates the chevron-shaped pattern imprinted on the beam density profile
by the OTSI in the cases of finite and infinite (i.e. with periodic boundary conditions along $x$) beam lengths. Galilean beam-frame coordinates $(\xi,\tau)=(v_bt-x,t)$ are used here, and the beam density maximum is located at $\xi\simeq 100 k_p^{-1}$ for $k_p \sigma_x \in (10,50)$. While the density modulations are uniform in the infinite beam case, they exhibit a clear spatial growth for finite $\sigma_x$. Figure~\ref{fig:fig1}(b) shows the 2D Fourier spectrum of the transverse $E_y$ fluctuations within a slice around the beam maximum, for $k_p\sigma_x=10$ [i.e. corresponding to the bottom plot of Fig.~\ref{fig:fig1}(a)] and $c\tau= 3338 k_p^{-1}$. A narrow continuum of modes located at $k_x \simeq k_p$ and $k_\perp \gtrsim k_p$ are excited, a characteristic feature of the OTSI \cite{Bret_POP_2010}.

The evolution of the RMS amplitude of the transverse $E_y$ and $B_z$ fields during the beam propagation in the plasma is presented in Fig.~\ref{fig:fig1}(c). In all cases considered, $E_y$ prevails over $B_z$, as is expected for the OTSI \cite{Bret_PRE_2010}. For an infinite beam, good agreement is found with
the temporal growth rate given by Eq.~\eqref{eq:G_OTSI}. By contrast, we observe a slowdown in the $E_y$ field growth as the beam length is decreased from $k_p\sigma_x=50$ to $k_p\sigma_x=10$.
To get a spatially resolved estimate of the effective OTSI growth rate in the finite-$\sigma_x$ simulations, we have fitted to an exponential the temporal evolution of the $E_y$ energy contained in the ``oblique'' spectral range $0.8 \le k_x/k_p \le 5$ and $0.8\le k_\perp/k_p \le 10$, and normalized the resulting growth rate, $\Gamma$, to $\Gamma_{\rm OTSI}$. In doing so, we have evaluated $k_\perp$ from the ratio of the $E_y$ and $E_x$ spectral energies integrated in the above $k$-range. Figure~\ref{fig:fig1}(d) displays the results of this procedure as a function of $k_p\sigma_x$. It is clearly seen that, even for $k_p\sigma_x \gg 1$, the effective growth rate is substantially smaller than is predicted for an unbounded system.

To understand the simulation results, we have developed a spatiotemporal model describing the growth of linear electrostatic oblique modes in a transversely homogeneous, relativistic  beam-plasma system, in the presence of immobile ions. The analysis is restricted to a 2D $(x,y)$ geometry, but it can be readily generalized to 3D. Let us denote the equilibrium quantities with a superscript $^{(0)}$, and perturbed variables with a superscript $^{(1)}$. Coupling the linearized, cold-fluid momentum and continuity equations for the beam (subscript $_b$) and plasma (subscript $_p$) electrons results in
\begin{gather}
\left(\partial_t +v_{b0}\partial_x\right)^2 n_b^{(1)} = n_{b0}\left( \gamma_{b0}^{-1} \partial_y^2 -\gamma_{b0}^{-3}\partial_x^2 \right)\phi^{(1)} \,,  \\
\partial_t^2 n_p^{(1)} = n_{p0}\left(\partial_y^2-\partial_x^2\right)\phi^{(1)} \,.    \label{eq:2}
\end{gather}
Next, using the linearized Poisson equation to express the perturbed electrostatic potential $\phi^{(1)}$ in terms of $n_{p}^{(1)}$ and $n_{b}^{(1)}$, one can obtain, after some algebra, the following differential equation for the perturbed plasma density
\begin{multline}
    \biggl[ \left(\partial_x^2+\partial_y^2\right)\left(\partial_t+v_{b0}\partial_x\right)^2\left(\partial_t^2+n_{p0}\right)   \\
      + \gamma_{b0}^{-1} n_{b0} \partial_y^2 \partial_t^2 \biggl] n_p^{(1)} = 0 \,,
    \label{eq:3}
\end{multline}
where the beam Lorentz factor has been supposed large enough that $\partial_y^2 \gg \gamma_{b0}^{-2} \partial_x^2$.
We now adopt the comoving coordinates defined above to express the plasma density perturbation as $n_p^{(1)} = \delta n_p(\tau,\xi) e^{-ik_p \xi + ik_\perp y}$, where $\delta n_p(\tau,\xi)$ represents a slowly varying envelope. Writing Eq.~\eqref{eq:3} in terms of the comoving variables and assuming that $k_p \gg v_{b0}^{-1} \partial_\tau, \partial_\xi$,
one can derive the following approximate differential equation satisfied by $\delta n_p$:
\begin{equation}
    \left(\partial_\tau^3 + v_b \partial_\tau^2 \partial_\xi + \frac{8i}{3^{3/2}} \Gamma_{\rm OTSI}^3 \right) \delta n_p = 0 \,.
    \label{eq:master_eq}
\end{equation}
This equation can be solved analytically for a semi-infinite electron (or electron-positron) beam whose front edge is located at $\xi=0$ (see Supplemental Material~\cite{SM}). Following Refs.~\cite{Decker_PoP_1996, Pathak_NJP_2015}, we assume an initial noise source throughout the beam, i.e., $\delta n_p (\tau=0,\xi) = \delta n_p(\tau,\xi=0) = S$, and  $\partial_\tau \delta n_p (\tau=0,\xi) = \partial^2_\tau \delta n_p( \tau=0,\xi) = 0$, where $S$ is some amplitude parameter. Such conditions mainly apply to a situation where the beam is created within the plasma or penetrates a plasma with a long density ramp. 
An asymptotic solution to Eq.~\eqref{eq:master_eq} can then be obtained in the $\tau \to \infty$ limit using a double Laplace transform and a saddle-point expansion~\cite{SM}. When $\xi \ll v_b \tau$, one finds
\begin{align}
    &\delta n_p (\tau,\xi) \simeq \frac{S}{\sqrt{6\pi}} \left(\frac{3\sqrt{3} v_b}{16\Gamma_{\rm OTSI}^3\,\xi \tau^2} \right)^{1/6} \nonumber \\
    &\times \exp\left[\frac{\sqrt{3}}{2^{2/3}} (\sqrt{3}+i)\Gamma_{\rm OTSI} \left(\frac{\xi}{v_b}\right)^{1/3} \tau^{2/3} - i \frac{\pi}{12} \right] \,.
    \label{eq:np_sp} 
\end{align}
This solution, similar to the asymptotic impulse solution of the TSI~\cite{Jones_POF_1983}, demonstrates the spatiotemporal character of the oblique instability. Different longitudinal $\xi$-slices of the beam experience different temporal evolutions, the fastest growth being present at the rear of the beam, as might be intuitively surmised. The same leading exponential term is found for an initial noise source localized at the beam front, as expected when the beam enters a sharp vacuum-plasma boundary~\cite{SM}. 

Further away from the beam front, i.e., for $\xi \ge v_b \tau$, the solution asymptotically evolves as
\begin{equation}
    \delta n_p(\tau,\xi) \simeq \frac{S}{3} \exp\left[\left(1+\frac{i}{\sqrt{3}}\right) \Gamma_{\rm OTSI}\,\tau \right]
    \label{eq:np_t} \,,
\end{equation}
which exhibits a purely temporal growth at the rate given by Eq.~\eqref{eq:G_OTSI}. In fact, the same exponential behavior sets in for $\xi \gtrsim v_b \tau/3$ but with a smaller prefactor~\cite{SM}.
In the comoving coordinates, the region of purely temporal growth recedes from the front to the rear of the beam at a velocity of $\sim v_b/3$.
Therefore, at a location $\xi$ behind the beam front, the instability initially grows in a purely temporal manner at a rate $\Gamma_{\rm OTSI}$, up to $\tau \simeq 3\xi v_b^{-1}$, after which spatiotemporal effects turn prominent and result in a slower growth. The same reasoning applied to a finite beam length $\sigma_x$ implies that for $\sigma_x \ll v_b \Gamma_{\rm OTSI}^{-1}$, the instability is essentially of spatiotemporal nature. The latter condition holds in particular for the short ultrarelativistic bunches produced in particle accelerators.

To support this analysis, we carried out 2D PIC simulations with a step-like beam profile. A neutral electron-positron ($e^-e^+$) pair beam was employed in order to avoid plasma wakefield excitation and minimize initial noise, and thus enable accurate comparison with the model (yet similar results were obtained with an electron beam~\cite{SM}). To reproduce even more closely the model assumptions, the beam entering the plasma was propagated ballistically till being completely immersed, and then (at $t=0$) let to evolve freely.
We used beam-plasma parameters relevant to FACET-II: $\gamma_b = 2\times 10^4$, $\alpha=(n_{b-}+n_{b+})/n_p = 0.06$ ($n_{b\pm}$ is the equal density of the beam electrons and positrons), and $n_p = 10^{20}\,\rm cm^{-3}$.
The simulation (moving) window covered the longitudinal range $-10 \le \xi \le \xi_{\rm max} = 150\,\rm \mu m$ (i.e. $-19 \le k_p \xi \le 282$), the beam front being placed at $\xi =0$. For these parameters, one finds
$\xi_{\rm max} <v_b \Gamma_{\rm OTSI}^{-1}$, hence the instability should evolve in a spatiotemporal manner.

Figure~\ref{fig:fig2}(a) displays (in solid curves) the spectral amplitude $\vert \widetilde{E}_y (k_x,k_\perp) \vert$ of the dominant oblique mode (at $k_x=k_p$ and $k_\perp \simeq 3k_p$) along the beam at different propagation distances $c\tau$, and in Fig.~\ref{fig:fig2}(b) the same quantity is plotted as a function of $c\tau$ for different positions $\xi$. Both figures show very good agreement with
the predicted spatiotemporal evolution $\propto \exp[(3/2^{2/3})\Gamma_{\rm OTSI} (\xi/c)^{1/3} \tau^{2/3}]$ of the instability (dashed lines). 

For large enough propagation distances ($c\tau \gtrsim 1000 k_p^{-1}$), the simulation curves in Fig.~\ref{fig:fig2}(a) peak at some position $\xi$, beyond which they rapidly decay. This behavior is due to the nonlinear saturation of the OTSI \cite{Thode_POF_1976, Sironi_APJ_2014}. 
The saturation mechanisms involved in the ultrarelativistic regime will be studied in a separate paper, yet one can exploit here their observed weak spatial dependence to further validate the theory. Indeed, assuming that the instability ceases when a certain field level is reached, the saturation position, $\xi_{\rm sat}$, should vary with $\tau$ as $\xi_\mathrm{sat} \propto \tau^{-2}$. This prediction matches well with the simulation results of Fig.~\ref{fig:fig2}(c), which plots $\xi_\mathrm{sat}$ vs. $\tau$.

Finally, to confirm the existence of a purely temporal regime, we repeated the same simulation but with a lower beam Lorentz factor ($\gamma_b = 20$), so that $\xi_{\rm max}> v_b \Gamma_{\rm OTSI}^{-1}$. Figure~\ref{fig:fig2}(d) shows that
the instability then grows at a rate that is essentially independent of the beam slice $\xi> 50\,\rm \mu m$. This nicely agrees with Eq.~\eqref{eq:np_t}, as shown by the dashed lines representing the predicted amplification of the initial (recorded at $c\tau=14k_p^
{-1}$) $\xi$-dependent fluctuations.

\begin{figure}[ht!]
\includegraphics[width=0.49\textwidth]{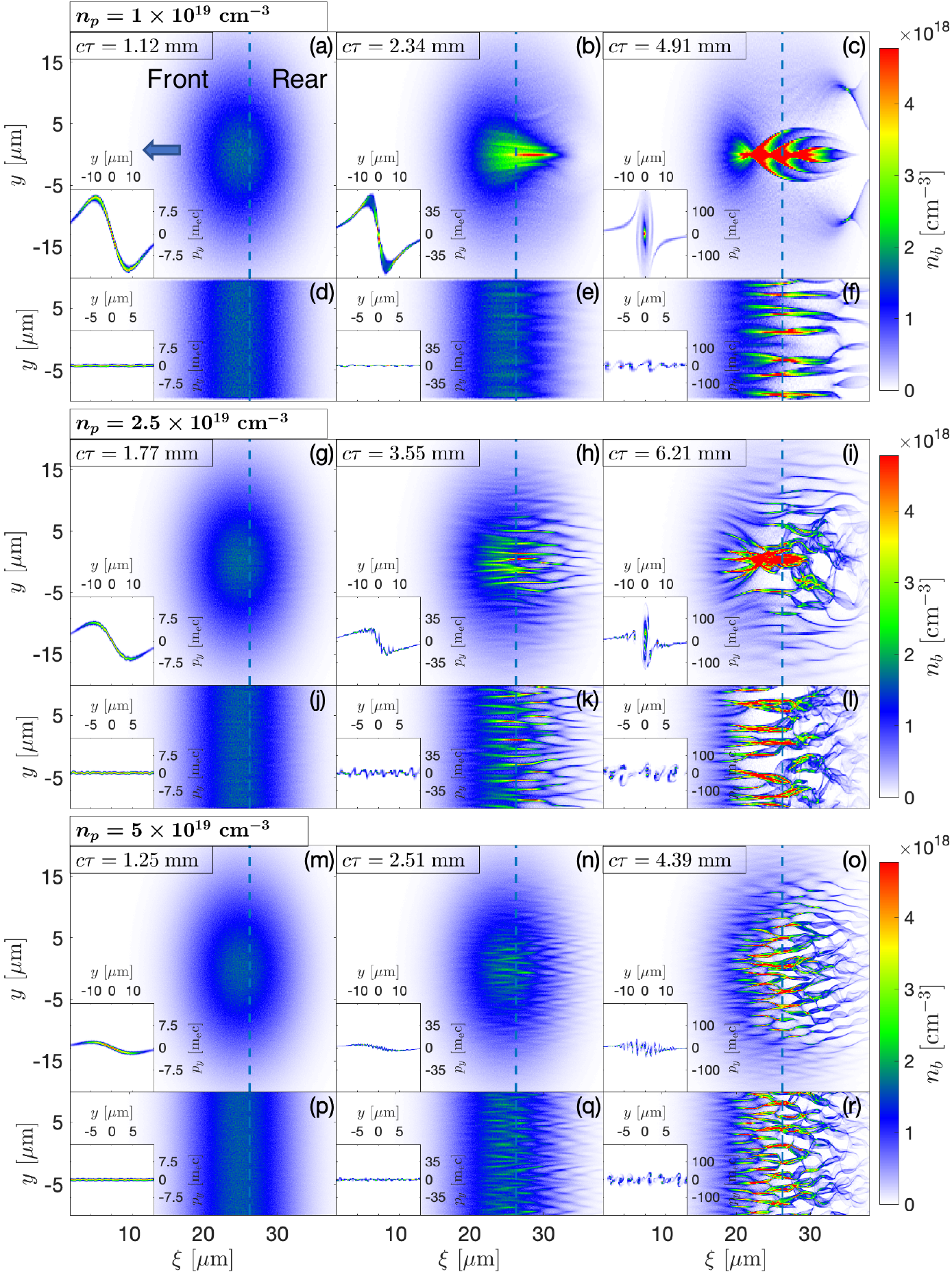}
\caption{\label{fig:fig3} Simulated electron beam density maps at different propagation distances
in a uniform plasma of density $n_p=\SI{1e19}{cm^{-3}}$ [(a)-(f)], $\SI{2.5e19}{cm^{-3}}$ [(g)-(l)], and $\SI{5e19}{cm^{-3}}$ [(m)-(r)]. The transverse beam profile is taken to be either finite with $\sigma_r=\SI{10}{\mu m}$ RMS width [(a)-(c), (g)-(i), and (m)-(o)] or infinite, i.e., with transverse periodic boundary conditions [(d)-(f), (j)-(l), and (p)-(r)]. In all cases, the beam has a 10~GeV energy ($\gamma_b = 2\times 10^4$), a Gaussian longitudinal profile with $\sigma_x=\SI{5}{\mu m}$ RMS length, a transverse normalized emittance $\epsilon_n=3\,\rm mm\,mrad$, and a peak density $n_b \simeq 1.5\times 10^{18} \rm cm^{-3}$ [i.e., $\alpha \simeq 0.15$ for (a)-(f),$\alpha \simeq 0.06$ for (g)-(l), and $\alpha \simeq 0.03$ for (m)-(r)], which would correspond to a total beam charge of $2\,\rm nC$ in 3D. The insets show the transverse beam phase space along the slices indicated by the dashed blue lines. }
\end{figure}

Another important finite-size effect is the excitation of plasma wakefields by nonneutral beams with relatively small transverse width ($\sigma_r$).
These fields act back on the beam to pinch it, which reinforces them and causes the beam to self-focus as it further propagates through the plasma~\cite{Corde_NC_2016}.
The time scale of beam self-focusing can be estimated by the inverse betatron frequency $\omega_\beta^{-1}=\sqrt{\gamma_b m_e/\partial_r W_\perp}$, where $W_\perp$ is the amplitude of  the transverse wakefield~\cite{Keinigs_POF_1987}. If this time scale is smaller than the effective growth time of the dominant oblique instability [i.e. lengthened by spatiotemporal effects, see Fig.~\ref{fig:fig1}(d)], the beam can shrink into a narrow and dense filament expelling the plasma electrons away from it, hence quenching the instability. For a beam with fixed charge and length, changing its transverse width affects both processes similarly, and so barely modifies their interplay. By contrast, raising the plasma density tends to favor the instability over the beam self-focusing.

We ran additional 2D PIC simulations to examine the interplay of the beam self-focusing and beam-plasma instability depending on the plasma density. Potential effects arising in a 3D geometry are discussed in the Supplemental Material~\cite{SM}. We considered a FACET-II-like electron beam (\SI{10}{GeV}, \SI{2}{nC}, $\sigma_x=\SI{5}{\mu m}$, $\sigma_r=\SI{10}{\mu m}$, peak density $n_b \simeq 1.5\times 10^{18}\,\rm cm^{-3}$, normalized emittance $\epsilon_n=3\,\rm mm\,mrad$) injected through a uniform plasma of different densities.
Each simulation was repeated with a transversely infinite (periodic) beam to suppress the effects of plasma wakefields and beam self-focusing.
Comparing $\omega_\beta^{-1}$ to the time scale of the spatiotemporal OTSI with the above parameters, one finds that beam self-focusing should dominate for $n_p \lesssim 10^{19}\,\rm cm^{-3}$ \cite{SM}. This prediction is confirmed by the simulation results depicted in Fig.~\ref{fig:fig3}. 
At $n_p = 10^{19}\,\rm cm^{-3}$ [Figs.~\ref{fig:fig3}(a)-(c)], the transverse wakefield starts focusing the finite-width beam [see its rotation in the transverse phase space in the inset of Fig.~\ref{fig:fig3}(a)] before the OTSI can impart significant modulations on the beam profile. This leads the whole beam to collapse into a narrow filament [Figs.~\ref{fig:fig3}(b)-(c)], hence inhibiting the OTSI in stark contrast with the equivalent infinite-beam simulation [Figs.~\ref{fig:fig3}(d)-(f)]. At $n_p=2.5\times 10^{19}\,\rm cm^{-3}$ [Figs.~\ref{fig:fig3}(g)-(i)], the self-focusing dynamics is slower, and so the competition between the two processes is more balanced. Still, although the OTSI-driven modulations have time to grow, a compressed filament eventually forms at the beam center [Fig.~\ref{fig:fig3}(i)], which is absent for an infinite beam width [Fig.~\ref{fig:fig3}(l)].
Finally, when further increasing the plasma density to $n_p = 5\times 10^{19}\,\rm cm^{-3}$ [Figs.~\ref{fig:fig3}(m)-(r)], the system dynamics is clearly governed by the OTSI, and, as expected, no significant difference arises when changing from a finite to an infinite beam width.

In conclusion, we have conducted the first spatiotemporal analysis of the oblique two-stream instability triggered by finite-size particle beams.
For ultrarelativistic, short-duration bunches, such as delivered by state-of-the-art particle accelerators, we have shown analytically that, in terms of the comoving coordinates $(\tau,\xi)$, the instability mainly evolves as a function of $(\xi/v_b)^{1/3}\tau^{2/3}$.
It develops from the head to the tail of the beam, and, within a fixed beam slice, more slowly than
in unbounded geometry. 
Close agreement has been found between the theory and PIC simulations in several beam-plasma setups.
Furthermore, when realistic finite-width electron beams are considered, self-focusing induced by plasma wakefields may hinder the instability growth, and thus dominate the beam dynamics. Neutral pair beams, though, can circumvent the limitation placed by wakefields, and facilitate laboratory investigations of ultrarelativistic streaming instabilities.
These results are critical to guide and interpret future experiments on high-energy beam-plasma interactions and their envisioned applications, such as the development of instability-based light sources.

\begin{acknowledgments}
This work was performed in the framework of the E-305 Collaboration. E-305 is a SLAC experiment which aims at the study of astrophysically relevant beam-plasma instabilities and at the generation of bright gamma rays. The work at CEA and LOA was supported by the ANR (UnRIP project, Grant No. ANR-20-CE30-0030). The work at LOA was also supported by the European Research Council (ERC) under the European Union's Horizon 2020 research and innovation programme (M-PAC project, Grant Agreement No. 715807). We acknowledge GENCI-TGCC for granting us access to the supercomputer IRENE under Grants No. 2019-A0060510786, 2020-A0080510786 and 2021-A0100510786 to run PIC simulations. J.~R.~Peterson was supported by DOE NNSA LRGF fellowship under grant DE-NA0003960. The work at SLAC was supported by U.S. DOE FES Grant No. FWP100331 and DOE Contract DE-AC02-76SF00515. UCLA was supported by U.S. Department of Energy Grant No. DE-SC001006 and NSF Grant No. 1734315.
\end{acknowledgments}

\providecommand{\noopsort}[1]{}\providecommand{\singleletter}[1]{#1}%
%
%\end{comment}

\end{document}

\title{Supplemental Material for \\
``Spatiotemporal dynamics of ultrarelativistic beam-plasma instabilities''}

\author{P.~San Miguel Claveria}
\email{pablo.san-miguel-claveria@polytechnique.edu}
\affiliation{LOA, ENSTA Paris, CNRS, Ecole Polytechnique, Institut Polytechnique de Paris, 91762 Palaiseau, France}
\author{X.~Davoine}
\affiliation{CEA, DAM, DIF, 91297 Arpajon, France}
\affiliation{Universit\'{e} Paris-Saclay, CEA, LMCE, 91680 Bruy\`{e}res-le-Ch\^{a}tel, France}
\author{J.~R.~Peterson}
\affiliation{SLAC National Accelerator Laboratory, Menlo Park, CA 94025, USA}
\affiliation{Stanford University, Physics Department, Stanford, CA 94305, USA}
\author{M.~Gilljohann}
\affiliation{LOA, ENSTA Paris, CNRS, Ecole Polytechnique, Institut Polytechnique de Paris, 91762 Palaiseau, France}
\author{I.~Andriyash}
\affiliation{LOA, ENSTA Paris, CNRS, Ecole Polytechnique, Institut Polytechnique de Paris, 91762 Palaiseau, France}
\author{R.~Ariniello}
\affiliation{University of Colorado Boulder, Department of Physics, Center for Integrated Plasma Studies, Boulder, Colorado 80309, USA}
\author{H.~Ekerfelt}
\affiliation{SLAC National Accelerator Laboratory, Menlo Park, CA 94025, USA}
\author{C.~Emma}
\affiliation{SLAC National Accelerator Laboratory, Menlo Park, CA 94025, USA}
\author{J.~Faure}
\affiliation{CEA, DAM, DIF, 91297 Arpajon, France}
\affiliation{Universit\'{e} Paris-Saclay, CEA, LMCE, 91680 Bruy\`{e}res-le-Ch\^{a}tel, France}
\author{S.~Gessner}
\affiliation{SLAC National Accelerator Laboratory, Menlo Park, CA 94025, USA}
\author{M.~Hogan}
\affiliation{SLAC National Accelerator Laboratory, Menlo Park, CA 94025, USA}
\author{C.~Joshi}
\affiliation{University of California Los Angeles, Los Angeles, CA 90095, USA}
\author{C.~H.~Keitel}
\affiliation{Max-Planck-Institut f\"ur Kernphysik, Saupfercheckweg 1, D-69117 Heidelberg, Germany}
\author{A.~Knetsch}
\affiliation{LOA, ENSTA Paris, CNRS, Ecole Polytechnique, Institut Polytechnique de Paris, 91762 Palaiseau, France}
\author{O.~Kononenko}
\affiliation{LOA, ENSTA Paris, CNRS, Ecole Polytechnique, Institut Polytechnique de Paris, 91762 Palaiseau, France}
\author{M.~Litos}
\affiliation{University of Colorado Boulder, Department of Physics, Center for Integrated Plasma Studies, Boulder, Colorado 80309, USA}
\author{Y.~Mankovska}
\affiliation{LOA, ENSTA Paris, CNRS, Ecole Polytechnique, Institut Polytechnique de Paris, 91762 Palaiseau, France}
\author{K.~Marsh}
\affiliation{University of California Los Angeles, Los Angeles, CA 90095, USA}
\author{A.~Matheron}
\affiliation{LOA, ENSTA Paris, CNRS, Ecole Polytechnique, Institut Polytechnique de Paris, 91762 Palaiseau, France}
\author{Z.~Nie}
\affiliation{University of California Los Angeles, Los Angeles, CA 90095, USA}
\author{B.~O'Shea}
\affiliation{SLAC National Accelerator Laboratory, Menlo Park, CA 94025, USA}
\author{D.~Storey}
\affiliation{SLAC National Accelerator Laboratory, Menlo Park, CA 94025, USA}
\author{N.~Vafaei-Najafabadi}
\affiliation{Stony Brook University, Stony Brook, NY 11794, USA}
\author{Y.~Wu}
\affiliation{University of California Los Angeles, Los Angeles, CA 90095, USA}
\author{X.~Xu}
\affiliation{SLAC National Accelerator Laboratory, Menlo Park, CA 94025, USA}
\author{J.~Yan}
\affiliation{Stony Brook University, Stony Brook, NY 11794, USA}
\author{C.~Zhang}
\affiliation{University of California Los Angeles, Los Angeles, CA 90095, USA}
\author{M.~Tamburini}
\affiliation{Max-Planck-Institut f\"ur Kernphysik, Saupfercheckweg 1, D-69117 Heidelberg, Germany}
\author{F.~Fiuza}
\affiliation{SLAC National Accelerator Laboratory, Menlo Park, CA 94025, USA}
\author{L.~Gremillet}
\email{laurent.gremillet@cea.fr}
\affiliation{CEA, DAM, DIF, 91297 Arpajon, France}
\affiliation{Universit\'{e} Paris-Saclay, CEA, LMCE, 91680 Bruy\`{e}res-le-Ch\^{a}tel, France}
\author{S.~Corde}
\email{sebastien.corde@polytechnique.edu}
\affiliation{LOA, ENSTA Paris, CNRS, Ecole Polytechnique, Institut Polytechnique de Paris, 91762 Palaiseau, France}

\date{\today}

\maketitle

\section{Spatiotemporal linear theory of the oblique two-stream instability}

\subsection{Derivation of the master equation}

Let us consider a transversely uniform, relativistic electron beam moving through an unmagnetized  electron-ion plasma. We seek to derive the equation governing the spatiotemporal perturbative evolution of the oblique two-stream instability (OTSI) as triggered by disturbances at the leading edge of the beam and within its body. According to previous studies \cite{Thode_POF_1976, Bret_PRE_2010}, the OTSI is mainly of electrostatic nature, so that the problem can be analytically addressed by combining the fluid conservation equations for the beam and plasma electrons and Poisson's equation. The ions will be assumed immobile throughout. 

In the following, the subscripts $_b$ and $_p$ will refer to the beam and plasma electrons, respectively. For population $\alpha =(b,p)$, $n_\alpha$ is the number density, $\gamma_\alpha$ the Lorentz factor, and $\mathbf{v_\alpha}$ the velocity. Moreover, $\phi$ will represent the electrostatic field potential. All plasma and field quantities will be linearized as $X(\mathbf{r},t)=X_0+X^{(1)}(\textbf{r},t)=X_0+\delta X(x,t)\exp[ i(\mathbf{k \cdot r}-\omega t)]$, where $X_0$ and $X^{(1)}$ denote unperturbed and perturbed quantities, and $\delta X$ represents the spatiotemporal envelope of the perturbation, characterized by its real wavenumber $\mathbf{k}$. The (longitudinal) $x$-axis is defined as the beam propagation direction. Only the case of an ultrarelavistic ($\gamma_{b0}\gg 1$), dilute ($n_{b0}/n_{p0}\ll 1$) beam will be treated and thermal effects will be neglected. 
Our analysis will be restricted to a 2D $(x,y)$ geometry, so that $\mathbf{k} = (k_x,k_\perp)$ and  $\mathbf{v_\alpha}=(v_{\alpha x},v_{\alpha y})$. %(note that the index $_y$ is noted $_\perp$ in the main text).
We will use units such that $e=m_e=c=\epsilon_0 =1.$ 

From the above assumptions, the momentum and continuity equations for the beam electrons can be written as
\begin{align}
    &(\partial_t+v_{b0} \partial_x)
    \begin{pmatrix}
    \gamma_{b0}^3v_{bx}^{(1)}\\
    \gamma_{b0}v_{by}^{(1)}
    \end{pmatrix}
    =
        \begin{pmatrix}
    \partial_x\phi^{(1)} \\
    ik_\perp \phi^{(1)}
    \end{pmatrix} \,, \\
&\partial_t n_b^{(1)}+\partial_x(n_bv_{bx})^{(1)}+ i k_\perp (n_b v_{by})^{(1)} = 0 \,.
\end{align}
Combining both equations leads to 
\begin{equation}
\left(\partial_t +v_{b0}\partial_x\right)^2 n_b^{(1)} = n_{b0}\left( \gamma_{b0}^{-1} k_\perp^2 -\gamma_{b0}^{-3}\partial_x^2 \right)\phi^{(1)} \,.
    \label{eq:S3}
\end{equation}

The plasma electrons are taken to be initially at rest ($v_{p0}=0$). Since $n_{b0}/n_{p0} \ll 1$ is further assumed, their dynamics can be treated nonrelativistically. As a result, they fulfill an equation similar to Eq.~\eqref{eq:S3} except for the changes $v_{b0} \to 0$ and $\gamma_{b0} \to 1$:
\begin{equation}
    \partial_t^2 n_p^{(1)} = n_{p0}\left(k_\perp^2-\partial_x^2\right)\phi^{(1)} \,.
    \label{eq:S4}
\end{equation}

We now plug Eq.~\eqref{eq:S4} into Poisson's equation 
\begin{equation}
\left(\partial_x^2-k_\perp^2\right)\phi^{(1)}=n_p^{(1)} + n_b^{(1)} \,,
\end{equation}
to obtain 
\begin{equation}
    \left( \partial_t^2+n_{p0}\right)n_p^{(1)}+n_{p0}n_b^{(1)}=0 \,.
    \label{eq:S5}
\end{equation}
Applying the $(\partial_x^2 -k_\perp^2)$ operator to  Eq.~\eqref{eq:S3} and using Eq.~\eqref{eq:S4}, we find
\begin{equation}
    \left(\partial_x^2-k_\perp^2\right)\left(\partial_t+v_{b0}\partial_x\right)^2n_b^{(1)}=\frac{n_{b0}}{n_{p0}}\left(\gamma_{b0}^{-3}\partial_x^2-\gamma_{b0}^{-1}k_\perp^2\right)\partial_t^2n_p^{(1)} \,.
    \label{eq:S6}
\end{equation}
Substituting this relation in Eq.~\eqref{eq:S5} gives the general equation verified by $n_p^{(1)}$:
\begin{equation}
    \left[\left(\partial_x^2-k_\perp^2\right)\left(\partial_t+v_{b0}\partial_x\right)^2\left(\partial_t^2+n_{p0}\right)+n_{b0}\left(\gamma_{b0}^{-3}\partial_x^2-\gamma_{b0}^{-1}k_\perp^2\right)\partial_t^2\right]n_p^{(1)} = 0 \,.
    \label{eq:S7}
\end{equation}

When discarding spatiotemporal effects, i.e., by taking $\partial_x = ik_x$ and $\partial_t = - i\omega$ ($\omega$ is the complex growth rate), the above equation reduces to the standard, electrostatic cold-fluid dispersion of the OTSI \cite{Thode_POF_1976}:
\begin{equation}
    1-\frac{n_{p0}}{\omega^2}-\frac{n_{b0}\left(\gamma_{b0}^{-3}k_x^2-\gamma_{b0}^{-1}k_\perp^2\right)}{\left(k_x^2+k_\perp^2\right)\left(\omega-k_xv_{b0}\right)^2} = 0 \,. \label{eq:S8}
\end{equation}

To describe the spatiotemporal evolution of the perturbation, it is convenient to make 
a Galilean transformation to the beam-frame coordinates $(\xi,\tau) \equiv (v_bt-x,t)$, and write $n_p^{(1)}=\delta n_p(\tau,\xi)\exp(-ik_p \xi+ik_\perp y)$. Equation~\eqref{eq:S7} can therefore be recast as
\begin{align}
    &\left( 2ik_0\partial_\xi + k_p^2+k_\perp^2 \right)\partial_\tau^2\left[\left(\partial_\tau+v_{b0}\partial_\xi \right)^2 - 2ik_p v_{b0}\left(\partial_\tau + v_{b0}\partial_\xi\right) + n_{p0} - (k_p v_{b0})^2 \right] \delta n_p \nonumber \\
    &+ \frac{n_{b0}k_\perp^2}{\gamma_{b0}}\left(\partial_\tau+v_b\partial_\xi-ik_0 v_{b0} \right)^2 \delta n_p = 0 \,.
\end{align}
We assume $\gamma_{b0}$ to be large enough that $\gamma_{b0}^{-3} \partial_x^2$ can be neglected relative to $\gamma_{b0}^{-1} k_\perp^2$ in the second term of Eq.~\eqref{eq:S7}.

We now choose $k_p = \sqrt{n_{p0}}/v_{b0}$ ($\equiv \omega_{p0}/v_{b0}$ in physical units, where $\omega_{p0}$ is the background plasma frequency) and adopt the slow-varying-amplitude approximation: $v_{b0}^{-1} \partial_\tau, \partial_\xi \ll k_p \sim \sqrt{n_{p0}}$. The above equation can then be approximated as
\begin{equation}
    \left( k_p^2 + k_\perp^2 \right) \partial_\tau^2 \left( \partial_\tau + v_{b0}\partial_\xi \right)\delta n_p  - ik_\perp^2 \frac{n_{b0} \sqrt{n_{p0}} }{2\gamma_{b0}} \delta n_p = 0 \,.
    \label{eq:S9}
\end{equation}
Finally, introducing
\begin{equation}
    \Gamma_0 = \left(\frac{1}{2\gamma_{b0}} \frac{n_{b0}}{n_{p0}} \frac{k_\perp^2}{k_p^2 + k_\perp^2}\right)^{1/3} \omega_{p0} \,,
    \label{eq:G0}
\end{equation}
we recover Eq.~(2) of the main text:
\begin{equation}
    \left[ \partial_\tau^2\left( \partial_\tau + v_{b0} \partial_\xi \right) + i\Gamma_0^3 \right] \delta n_p = 0 \,.
    \label{eq:master_eq}
\end{equation}
Note that this equation could also have been obtained by applying the method of \cite{Rostomyan_POP_2000}
to the OTSI dispersion relation \eqref{eq:S8}. While quite straightforward, this method does not allow us to identify the physical quantity exactly governed by the resulting equation, hence the interest of our detailed derivation. It should also be pointed out that Eq.~\eqref{eq:master_eq} holds as well for an ultrarelativistic pair beam, in which case $n_{b0}$ in Eq.~\eqref{eq:G0} should be understood as the sum of the unperturbed number densities of the beam electrons and positrons.

\subsection{Analysis of the spatiotemporal behavior of the OTSI}

An important observation is that Eq.~\eqref{eq:master_eq} is formally identical to Eq.~(6) of \cite{Rostomyan_POP_2000} (assuming vanishing dissipation and group velocity of the unstable wave packet), which describes the longitudinal ($k_\perp=0$) two-stream instability (TSI).
In the following, we will solve Eq.~\eqref{eq:master_eq} for two sets of initial and boundary conditions. The first configuration is characterized by a Dirac function source at $\tau=\xi=0$, and will give the Green's function response of the system, recovering the well-known spatiotemporal impulse behavior of the TSI \cite{Bers_Handbook_1983, Jones_POF_1983, Rostomyan_POP_2000}. 
As an alternative setup, we will consider disturbances both throughout the beam ($\xi \ge 0$) at $\tau=0$ and at the beam front ($\xi=0$) at $\tau \ge 0$. Similar conditions were used in \cite{Decker_PoP_1996} in the case of Raman forward scattering of short laser pulses, and, more recently, to investigate the spatiotemporal properties of the current filamentation instability (CFI) \cite{Pathak_NJP_2015}. We will show that the solution to Eq.~\eqref{eq:master_eq} then exhibits both spatiotemporal and purely temporal behaviours.

Following \cite{Decker_PoP_1996}, we consider the double Laplace transform of $\delta n_p$, defined by
\begin{equation}
    \delta n_p(\alpha,\beta) = \int_0^\infty d\tau \int_0 ^\infty d \xi\, \delta n_p (\tau,\xi) e^{-i \alpha \tau - i\beta \xi} \,.
\end{equation}
To obtain the Green's function of the system, we put $\delta(\tau) \delta(\xi)$ in the r.h.s of Eq.~\eqref{eq:master_eq}, and apply the above transformation:
\begin{align}
    i \left( \alpha^3 + \alpha^2 \beta v_{b0} + \Gamma_0^3 \right) \delta n_p(\alpha,\beta) = -1 \,.
\end{align}
Inverting the Laplace transforms gives
\begin{equation}
    \delta n_p(\tau,\xi) = \frac{i}{4\pi^2 v_{b0}} \int_{C_\alpha} d\alpha\, e^{i\alpha \tau} \int_{C_\beta} d\beta \frac{e^{i \beta \xi} }{\alpha^2 \left( \beta + \frac{\alpha^3+\Gamma_0^3}{\alpha^2 v_{b0}} \right)}  \,.
    \label{eq:ILT1}
\end{equation}
with the Bromwich contours, $C_\alpha$ and $C_\beta$, extending from, respectively, $\Re\,\alpha = -\infty$ to $\Re\,\alpha = +\infty$ and $\Re\,\beta = -\infty$ to $\Re\,\beta = +\infty$, and lying below the integrand singularities. The integral over $\beta$ can be readily evaluated using the residue theorem for the pole at $\beta = - \frac{(\alpha^3+\Gamma_0^3)}{\alpha^2 v_{b0}}$:
\begin{equation}
    \int_{C_\beta} d\beta \frac{e^{i \beta \xi} }{ \beta + \frac{\alpha^3+\Gamma_0^3}{\alpha^2 v_{b0}}} =  2\pi i e^{-i\frac{(\alpha^3 + \Gamma_0^3)}{\alpha^2 v_{b0}}\xi} H(\xi) \,.
\end{equation}
The Heaviside function $H(\xi)$ expresses the fact that when $\xi < 0$, one can close the $C_\alpha$ contour with a semicircle of infinite radius in the half-plane $\Im (\alpha) < 0$, along which the exponential vanishes. Since no singularity is enclosed, the integral also vanishes.

Equation~\eqref{eq:ILT1} then becomes
\begin{equation}
    \delta n_p(\tau,\xi) = -\frac{1}{2\pi v_{b0}} \int_{C_\alpha} \frac{d\alpha}{\alpha^2} \, e^{i\alpha \tau - i\frac{(\alpha^3 + \Gamma_0^3)}{\alpha^2 v_{b0}}\xi} H(\xi) \,.
    \label{eq:inv_lap_1}
\end{equation}
This integral vanishes for $\tau < \xi/v_{b0}$. When $\tau > \xi/v_{b0}$, the residue of the essential singularity at $\alpha=0$ cannot be exactly calculated. To get an asymptotic approximation when $\tau \to \infty$, we use the saddle-point method \cite{Oughstun_Book_2009}. To this purpose, we introduce the dimensionless parameter $\theta = \xi/v_{b0} \tau$, and write the above expression as 
\begin{equation}
    \delta n_p(\tau,\xi) = -\frac{1}{2\pi v_{b0}} \int_{C_\alpha} d\alpha g(\alpha) e^{\tau h(\alpha,\theta)} \,,
    \label{eq:inv_lap_2}
\end{equation}
where
\begin{align}
    &h(\alpha,\theta) = i\left[(1-\theta)\alpha - \frac{\Gamma_0^3 \theta}{\alpha^2} \right] \nonumber \,, \label{eq:h} \\
    &g(\alpha) = \frac{1}{\alpha^2} \,,
\end{align}
The saddle points of $h(\alpha,\theta)$ are located at $\alpha_{\rm sp0}(\theta) = \Gamma_0 \left(\frac{2\theta}{1-\theta}\right)^{1/3} e^{i\pi}$ and $\alpha_{\rm sp \pm}(\theta) = \Gamma_0 \left(\frac{2\theta}{1-\theta}\right)^{1/3} e^{\pm i\pi/3}$, where $h(\alpha_{\rm sp 0}(\theta),\theta) = 3 \Gamma_0 \left(\frac{1-\theta}{2}\right)^{1/3} e^{-i\pi/2}$, $h(\alpha_{\rm sp +}(\theta),\theta) = 3 \Gamma_0 \left(\frac{1-\theta}{2}\right)^{1/3} e^{i 5\pi/6}$ and $h(\alpha_{\rm sp-}(\theta),\theta) = 3 \Gamma_0 \left(\frac{1-\theta}{2}\right)^{1/3} e^{i\pi/6}$.
The dominant saddle point is $\alpha_{\rm sp-}(\theta)$ since $\Re\, h(\alpha_{\rm sp-}(\theta),\theta)> \Re\, h(\alpha_{\rm sp0}(\theta),\theta), \Re\, h(\alpha_{\rm sp+}(\theta),\theta)$ for $0 < \theta < 1$, and the initial contour $C_\alpha$ can be continuously deformed to an Olver-type path with respect to $\alpha_{\rm sp-}$,  $C_{\rm sdp}$, along which $\Re\, h(\alpha, \theta) < \Re\, h(\alpha_{\rm sp-},\theta)$. Furthermore, this path goes from $(\Re\, \alpha, \Im\, \alpha)=(-\infty,+\infty)$ to $(\Re\, \alpha, \Im\, \alpha)=(+\infty,+\infty)$ and follows the path of steepest descent in the vicinity of $\alpha_{\rm sp-}$. Since $C_\alpha$ is deformable to $C_{\rm sdp}$ without crossing $\alpha = 0$ for $0 < \theta < 1$, there is no residue contribution of the pole singularity \cite{Oughstun_Book_2009}. The resulting integral can therefore be approximated as \cite{Oughstun_Book_2009}
\begin{equation}
    \int_{C_{\rm sdp}} d\alpha \, g(\alpha) e^{i \tau h(\alpha,\theta)} \simeq g(\alpha_{\rm sp-}) \sqrt{\frac{2\pi}{\tau \vert h''(\alpha_{\rm sp-},\theta) \vert}} e^{ \tau h(\alpha_{\rm sp-},\theta) + i\psi_{\rm sdp}} \,,
\end{equation}
where $h''(\alpha,\theta) \equiv \partial^2_\alpha h(\alpha,\theta)$ and
\begin{equation}
    \psi_{\rm sdp} = \frac{\pi}{2} - \frac{\mathrm{arg}\left[ h''(\alpha_{\rm sp-},\theta)\right]}{2}
\end{equation}
is the slope angle of the steepest-descent path through $\alpha_{\rm sp-}(\theta)$ \cite{Oughstun_Book_2009}. Using $g(\alpha_{\rm sp-}(\theta)) = \left( \frac{1-\theta}{2\theta} \right)^{2/3} \frac{e^{2i \pi/3}}{\Gamma_0}$ and substituting the above expressions for $h(\alpha_{\rm sp-}(\theta),\theta)$ and $h''(\alpha_{\rm sp-}(\theta),\theta)$, we find $\psi_{\rm sdp} = \pi/12$ and    
\begin{align}
    \delta n_p &\simeq  -\frac{1}{2\sqrt{3 \pi} \Gamma_0 v_{b0} \sqrt{\Gamma_0 \tau \theta}} \exp\left[ 3\Gamma_0 \tau \left(\frac{1-\theta}{2} \right)^{2/3} \theta^{1/3} e^{i\pi/6} + i\frac{3\pi}{4} \right]  \nonumber \\
    &\simeq \frac{1}{2\sqrt{3\pi} \Gamma_0 v_{b0} \sqrt{\Gamma_0 \xi/v_{b0}}} \exp\left[\frac{3}{2^{5/3}} (\sqrt{3}+i) \Gamma_0 \left( \tau -\frac{\xi}{v_{b0}}\right)^{2/3} \left(\frac{\xi}{v_{b0}}\right)^{1/3} - i\frac{\pi}{4} \right] \,,
\end{align}
for $0 < \xi/v_{b0} \tau < 1$, and $\delta n_p = 0$ otherwise. This solution coincides with that previously obtained for the TSI \cite{Bers_Handbook_1983, Jones_POF_1983,Rostomyan_POP_2000}, albeit with a different expression for $\Gamma_0$. It describes a wave packet growing approximately as
\begin{equation}
    \delta n_p \propto \exp\left[\frac{3^{3/2}}{2^{5/3}} \Gamma_0 \left( \tau -\frac{\xi}{v_{b0}}\right)^{2/3} \left(\frac{\xi}{v_{b0}}\right)^{1/3} \right]
    = \exp\left[\frac{3^{3/2}}{2^{5/3}} \Gamma_0 \left(\frac{x}{v_{b0}}\right)^{2/3} \left(t-\frac{x}{v_{b0}}\right)^{1/3} \right]
\end{equation}
 while propagating through the plasma. The peak of the wave packet is located at $\xi=v_{b0}\tau/3$ (i.e. $\theta = 1/3$), and so moves at a speed of $2v_{b0}/3$ in the laboratory. Its amplitude grows with time at a rate
\begin{equation}
    \frac{\sqrt{3}}{2}\Gamma_0 = \frac{\sqrt{3}}{2^{4/3}} \left( \frac{1}{\gamma_{b0}} \frac{n_{b0}}{n_{p0}} \frac{k_\perp^2}{k_p^2 + k_\perp^2} \right)^{1/3} \omega_{p0} \,,
    \label{eq:G_max}
\end{equation} 
that is, the well-known temporal growth rate of the cold-fluid OTSI in the $\gamma_{b0} \gg 1$ limit \cite{Ferch_PP_1975, Bret_PRE_2010}, as given by Eq.~(1) of the main text.

The above impulsive solution should be of most significance when the relativistic beam crosses a sharp vacuum-plasma boundary. The initial disturbance is then localized at $\xi=0$ given the absence of plasma at $\xi > 0$. This solution, however, is likely inappropriate if the uniform plasma region where the instability mainly develops is preceded by a long increasing density ramp, or if the beam is directly created within the plasma. Such situations can be modeled by the following initial and boundary conditions \cite{Decker_PoP_1996, Pathak_NJP_2015}:
\begin{align}
    &\delta n_p (\tau=0,\xi) = \delta n_p(\tau,\xi=0) = S \,, \\
    &\partial_\tau \delta n_p (\tau=0,\xi) = \partial^2_\tau \delta n_p (\tau=0,\xi) = 0 \,,
    \label{eq:IBC}
\end{align}
where $S$ is some initial fluctuation amplitude. The double Laplace transformation of Eq.~\eqref{eq:master_eq} then yields
\begin{equation}
     \left( \alpha^3 + \alpha^2 \beta v_{b0} + \Gamma_0^3 \right) \delta n_p(\alpha,\beta) = - i \frac{\alpha}{\beta} (\beta v_{b0} + \alpha) S \,,
\end{equation}
use being made of $\delta n_p (\tau=0,\beta) = -i S/\beta$, $\delta n_p (\alpha,\xi=0) = -iS/\alpha$, and $\partial_\tau \delta n_p(\tau=0,\beta) = \partial^2_\tau \delta n_p (\tau=0,\beta)= 0$.
It follows that
\begin{equation}
    \delta n_p (\alpha,\beta) =  - \frac{(\alpha + \beta v_{b0})S }{v_{b0} \alpha \beta \left( \beta + \frac{\alpha^3 + \Gamma_0^3}{\alpha^2 v_{b0}} \right)}  \,,
    \label{eq:DLT2}
\end{equation}
and hence
\begin{equation}
    \delta n_p(\tau,\xi) = -\frac{S}{4\pi^2 v_{b0}} \int_{C_\alpha} d\alpha \, e^{i\alpha \tau} \int_{C_\beta} d\beta \, \frac{\left(\alpha + \beta v_{b0} \right) e^{i \beta \xi} }{\alpha \beta \left( \beta + \frac{\alpha^3 + \Gamma_0^3}{\alpha^2 v_{b0}} \right)}  \,,
\end{equation}
where $C_\alpha$ and $C_\beta$ denote again the Bromwich contours. We recast this double integral as 
\begin{equation}
    \delta n_p(\tau,\xi) =  - \frac{S}{2\pi} \int_{C_\alpha} d\alpha \, I(\alpha,\xi) e^{i\alpha \tau} \,,
    \label{eq:ILT2}
\end{equation} 
where
\begin{equation}
    I(\alpha,\xi) = \frac{1}{2\pi v_{b0}} \int_{C_\beta} d\beta \, \frac{\left(\alpha + \beta v_{b0} \right) e^{i \beta \xi} }{\alpha \beta \left( \beta + \frac{\alpha^3+\Gamma_0^3}{\alpha^2 v_{b0}} \right)}  \,.
\end{equation}
Applying the residue theorem to the pole singularities at $\beta = 0$ and $\beta =  - \frac{(\alpha^3 + \Gamma_0^3)}{\alpha^2 v_{b0}}$ gives
\begin{equation}
    I(\alpha,\xi) = i \left[ \frac{\alpha^2}{\alpha^3 + \Gamma_0^3} + \frac{\Gamma_0^3}{\alpha(\alpha^3+\Gamma_0^3)} e^{-i\frac{(\alpha^3 + \Gamma_0^3)}{\alpha^2 v_{b0}}\xi} \right] H(\xi) \,.
    \label{eq:ILT3}
\end{equation}
Plugging Eq.~\eqref{eq:ILT3} into Eq.~\eqref{eq:ILT2} yields 
\begin{equation}
    \delta n_p(\tau,\xi) = - i \frac{S}{2\pi} \int_{C_\alpha} d\alpha \, e^{i\alpha \tau} \left[ \frac{\alpha^2}{\alpha^3+\Gamma_0^3} + \frac{\Gamma_0^3}{\alpha (\alpha^3+\Gamma_0^3)} e^{-i\frac{(\alpha^3 + \Gamma_0^3)}{\alpha^2 v_{b0}}\xi}  \right] H(\xi) \,,
\end{equation}
which can be expressed in the form 
\begin{equation}
    \delta n_p(\tau,\xi) = S \left[ J_1(\tau,\xi) + J_2(\tau,\xi) \right] H(\xi) \,,
    \label{eq:J1_J2}
\end{equation} 
where
\begin{align}
    &J_1(\tau) = - \frac{i}{2\pi} \int_{C_\alpha} d\alpha \, \frac{\alpha^2 e^{i\alpha \tau} }{\alpha^3+\Gamma_0^3}  \,, \label{eq:J1_1} \\
    &J_2(\tau,\xi) = - \frac{i \Gamma_0^3}{2\pi} \int_{C_\alpha} d\alpha \, \frac{e^{i \alpha \tau -i\frac{(\alpha^3 + \Gamma_0^3)}{\alpha^2 v_{b0}}\xi}}{\alpha (\alpha^3+\Gamma_0^3)} \,. \label{eq:J2_1}
\end{align}

Let us first address $J_1$. For the same reason as above, $J_1(\tau<0) = 0$, as expected from causality. Its integrand, $G_1(\alpha)$, has simple poles at $\alpha_{\rm p0} = -\Gamma_0$ and $\alpha_{\rm p\pm} = \Gamma_0 e^{\pm i \pi/3}$, and their residues are
\begin{align}
    &\mathrm{Res}\left[G_1(\alpha); \alpha_{\rm p 0} \right] = \frac{1}{3} e^{-i \Gamma_0 \tau} \,, \\
    &\mathrm{Res}\left[G_1(\alpha); \alpha_{\rm p \pm}\right] = \frac{1}{3} e^{\frac{i}{2} (1 \pm i \sqrt{3}) \Gamma_0 \tau} \,.        
\end{align}
Summing the residues leads to
\begin{equation}
    J_1(\tau) = \frac{1}{3}\left[ e^{-i \Gamma_0 \tau} + 2 \cosh \left( \frac{\sqrt{3}}{2} \Gamma_0 \tau\right) e^{i \frac{\Gamma_0}{2} \tau} \right] H(\tau) \,.
    \label{eq:J1_2}
\end{equation} 

Similarly to Eq.~\eqref{eq:inv_lap_2}, the integral $J_2$ cannot be exactly evaluated, and so instead we approximate it in the $\tau \to \infty$ limit. We therefore rewrite $J_2$ as
\begin{equation}
    J_2(\tau,\xi) = -\frac{i \Gamma_0^3}{2\pi} \int_{C_\alpha} d\alpha \, g(\alpha) e^{\tau h(\alpha,\theta)} \,, 
    \label{eq:J2_2}
\end{equation}
where $h(\alpha,\theta)$ is defined by Eq.~\eqref{eq:h} and
\begin{equation}
    g(\alpha) = \frac{1}{\alpha (\alpha +\Gamma_0^3)} \,. \label{eq:g}
\end{equation}
The function $g(\alpha)$ possesses four simple pole singularities at $\alpha = 0$, $\alpha_{\rm p 0}$ and $\alpha_{\rm p \pm}$ (as defined above). We note that the saddle points of $h(\alpha,\theta)$ fulfill $\mathrm{arg}\left[\alpha_{\rm sp 0}(\theta)\right] = \mathrm{arg}\left[\alpha_{\rm p 0}\right]$ and $\mathrm{arg}\left[\alpha_{\rm sp \pm}(\theta)\right] = \mathrm{arg}\left[\alpha_{\rm p \pm}\right]$. Moreover, $\alpha_{\rm sp 0}(\theta) \to \alpha_{\rm p0}$ and $\alpha_{\rm sp \pm}(\theta) \to \alpha_{\rm p\pm}$ when $\theta \to 1/3$.

The asymptotic expansion of the integral \eqref{eq:J2_2} is more involved than that of \eqref{eq:inv_lap_2}, owing to the interaction between the dominant saddle point $\alpha_{\rm sp-}(\theta)$ of $h(\alpha,\theta)$ and the pole singularity at $\alpha_{\rm p-} = \Gamma_0 e^{-i \pi/3}$ when the parameter $\theta$ varies over $(0,1)$. Specifically, for $\theta \le 1/3$, the path of steepest descent through $\alpha_{\rm sp-}(\theta)$ passes above $\alpha_{\rm p-}$ so that the residue contribution of the latter should be taken into account. For $\theta > 1/3$, by contrast, the path of steepest descent through $\alpha_{\rm sp-}(\theta)$ passes below $\alpha_{\rm p-}$, so that there is no residue contribution. As a result,
\begin{equation}
    \int_{C_\alpha} d\alpha \, g(\alpha) e^{\tau h(\alpha,\theta)} = \int_{C_{\rm sdp}} d\alpha \, g(\alpha) e^{\tau h(\alpha,\theta)} + 
    \left\{
    \begin{array}{ll}
    i 2\pi\,\mathrm{Res}\left[ g(\alpha); \alpha_{\rm p-} \right] e^{\tau h(\alpha_p,\theta)}  &\theta < 1/3 \\
     i \pi \,\mathrm{Res}\left[ g(\alpha);\alpha_{\rm p-} \right] e^{\tau h(\alpha_p,\theta)}  &\theta = 1/3 \\
     0  &\theta > 1/3 \,,
    \end{array}
  \right.
  \label{eq:J2_3}
\end{equation}
where $C_{\rm sdp}$ is an Olver-type path with respect to $\alpha_{\rm sp-}$, as defined above \cite{Oughstun_Book_2009}.

Another difficulty arises when the saddle point $\alpha_{\rm sp-}(\theta)$ approaches the pole at $\alpha_{\rm p-}$, so that $\vert h(\alpha_{\rm sp-}(\theta),\theta) - h(\alpha_{\rm p-},\theta) \vert \tau$ may not be necessarily large. Introducing 
\begin{equation}
    \Delta (\theta) = \left[ h(\alpha_{\rm sp-}(\theta),\theta) - h(\alpha_{\rm p-},\theta) \right]^{1/2} \,,
    \label{eq:Delta}
\end{equation}
a uniform asymptotic approximation of the saddle-point integral is then given by~\cite{Oughstun_Book_2009}
\begin{align}
    \int_{C_{\rm sdp}} d\alpha \, g(\alpha) e^{\tau h(\alpha,\theta)} \simeq e^{\tau h(\alpha_{\rm sp-},\theta)} \Bigg\{&\pm i \mathrm{Res} \left[g(\alpha);\alpha_{\rm p-}\right]\, \mathrm{erfc} \left[\mp \sqrt{\tau} \Delta(\theta)\right] e^{\tau[ h(\alpha_{\rm sp-},\theta) -h(\alpha_{\rm p-},\theta)]} \nonumber \\
    &+ \sqrt{\frac{\pi}{\tau}} \left[ g(\alpha_{\rm sp-}) \sqrt{\frac{2}{\vert h''(\alpha_{\rm sp-},\theta) \vert}}  e^{i \psi_{\rm sdp}} + \frac{\mathrm{Res} \left[g(\alpha);\alpha_{\rm p-}\right]}{\Delta(\theta)}  \right] \Bigg\} \,,
    \label{eq:I_SDP_pole1}
\end{align}
when $\Im \Delta (\theta) \gtrless 0$, and by
\begin{equation}
    \int_{C_{\rm sdp}} d\alpha \, g(\alpha) e^{\tau h(\alpha,\theta)} \simeq \sqrt{\frac{2 \pi}{\tau \vert h''(\alpha_{\rm sp-},\theta) \vert }} e^{\tau h(\alpha_{\rm sp-},\theta) + i \psi_{\rm sdp}} \left[ g(\alpha_{\rm sp-}) - \frac{\mathrm{Res} \left[g(\alpha);\alpha_{\rm p-}\right]}{\alpha_{\rm sp-} - \alpha_{\rm p-}} - \mathrm{Res} \left[g(\alpha);\alpha_{\rm p-}\right] \frac{h^{(3)}(\alpha_{\rm sp-})}{6h''(\alpha_{\rm sp-})} \right]
    \label{eq:I_SDP_pole2}
\end{equation}
when $\Delta (\theta) = 0$.

The function $\mathrm{erfc}(z) = (2/\sqrt{\pi}) \int_z^\infty du\, e^{-u^2}$ is the complementary error function. In Eq.~\eqref{eq:Delta}, the argument of $\Delta (\theta)$ is defined so that \[ \lim_{\alpha_{\rm p-} \to \alpha_{\rm sp-}} \mathrm{arg}\left[ \Delta (\theta) \right] = \mathrm{arg}\,(\alpha_{\rm p-} - \alpha_{\rm sp-}) - \psi_{\rm sdp} + 2\pi n \,,\] the integer $n$ being chosen so that $\mathrm{arg}\,\left[\Delta(\theta) \right]$ lies within $(-\pi,\pi)$. Here, we have $\psi_{\rm sdp} = \pi/12$ and $\mathrm{arg}\,(\alpha_{\rm p-} - \alpha_{\rm sp-}) = -\pi/3$ (resp. $2\pi/3$) for $\theta < 1/3$ (resp. $>1/3$). Hence, $\mathrm{arg}\,\left[\Delta (\theta)\right] = -5\pi/12$ (resp. $7\pi/12$) for $\theta < 1/3$ (resp. $>1/3$).

Combining Eqs.~\eqref{eq:J1_J2}, \eqref{eq:J2_3} and \eqref{eq:I_SDP_pole1}, and using the expressions
\begin{align}
    &\mathrm{Res}\left[g(\alpha);\alpha_{\rm p-}\right] = -\frac{1}{3\Gamma_0^3} \,, \nonumber \\
    &h(\alpha_{\rm p-},\theta) = \Gamma_0 e^{i \pi/6} \,, \nonumber \\
    &\vert \Delta (\theta) \vert = \left[ \Gamma_0 \left( 1- 3(\frac{1-\theta}{2})^{2/3} \theta^{1/3} \right) \right]^{1/2} \,, \nonumber \\
    &h^{(3)}(\alpha_{\rm sp-},\theta=1/3) = \frac{8 e^{i \pi/6}}{\Gamma_0^2} \,, \nonumber \\
    & \lim_{\alpha \to \alpha_{\rm p-}} \left\{ g(\alpha) - \frac{\mathrm{Res}\left[g(\alpha);\alpha_{\rm sp-}\right]}{\alpha-\alpha_{\rm p-}} \right\} = \frac{2e^{i\frac{\pi}{3}}}{3\Gamma_0^4} \,, \nonumber 
\end{align}
one can obtain the following approximate solution for $\tau \to \infty$:
\begin{align}
    \delta n_p(\tau,\xi) &\simeq \frac{S}{6} e^{3 \Gamma_0 \tau \left( \frac{1-\theta}{2} \right)^{2/3} \theta^{1/3}e^{i\pi/6} - i\pi/12} \Bigg\{ \frac{1}{\sqrt{\pi \Gamma_0 \tau}} \left[ 2^{1/3}\sqrt{3}\frac{(1-\theta)^{2/3}}{\theta^{1/6} (1-3\theta)} - \frac{1}{\left[1-3(\frac{1-\theta}{2})^{2/3} \theta^{1/3} \right]^{1/2}}\right] \nonumber \\
    &+ \mathrm{erfc}\,\left[\sqrt{\tau} \vert \Delta (\theta) \vert e^{i\frac{\pi}{12}} \right] e^{\Gamma_0 \tau [1- 3(\frac{1-\theta}{2})^{2/3} \theta^{1/3}] e^{i\pi/6} + i\pi/12 } \Bigg\} \,, \quad 0 < \theta < 1/3 \,,  \label{eq:np_final_1} \\
    \delta n_p(\tau,\xi) &\simeq \frac{S}{6}\left(1 + \frac{4}{3\sqrt{\pi \Gamma_0 \tau}} \right) e^{\Gamma_0 \tau e^{i\pi/6}} \,, \quad \theta = 1/3 \,.  \label{eq:np_final_2} \\
    \delta n_p(\tau,\xi) &\simeq \frac{S}{6} e^{3 \Gamma_0 \tau \left( \frac{1-\theta}{2} \right)^{2/3} \theta^{1/3}e^{i\pi/6} - i\pi/12} \Bigg\{ \frac{1}{\sqrt{\pi \Gamma_0 \tau}} \left[ 2^{1/3}\sqrt{3}\frac{(1-\theta)^{2/3}}{\theta^{1/6} (1-3\theta)} + \frac{1}{\left[1-3(\frac{1-\theta}{2})^{2/3} \theta^{1/3} \right]^{1/2}}\right] \nonumber \\
    &+ \mathrm{erfc}\,\left[\sqrt{\tau} \vert \Delta (\theta) \vert e^{i\frac{\pi}{12}} \right] e^{\Gamma_0 \tau [1- 3(\frac{1-\theta}{2})^{2/3} \theta^{1/3}] e^{i\pi/6} + i\pi/12 } \Bigg\} + \frac{S}{3} e^{\Gamma_0 \tau e^{i\pi/6}} \,, \quad 1/3 < \theta < 1 \,, \\
    \delta n_p(\tau,\xi) &\simeq \frac{S}{3} e^{\Gamma_0 \tau e^{i\pi/6}} \,,  \quad \theta > 1 \,. \label{eq:np_final_3}
\end{align}

\begin{figure}
    \centering
    \includegraphics[width=0.4\textwidth]{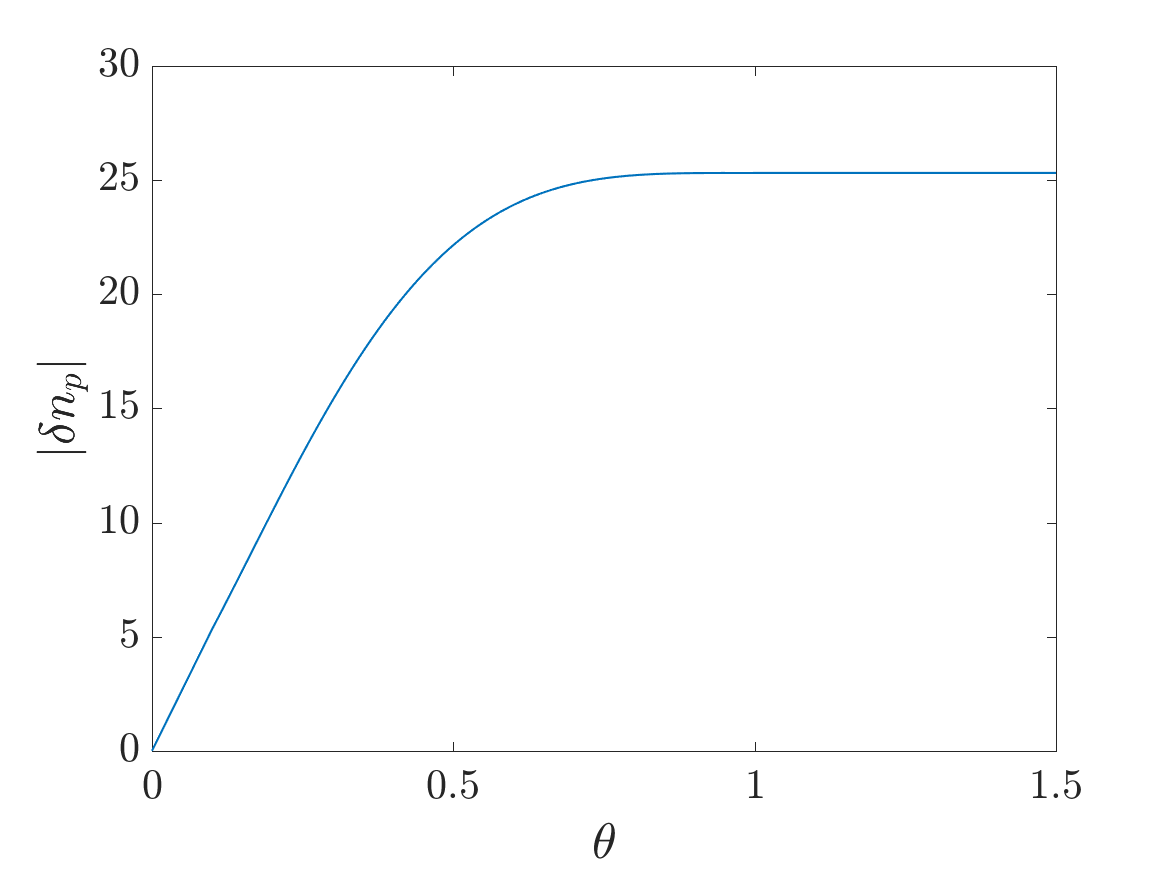}
    \caption{Asymptotic profile of the OTSI as a function of $\theta = \xi/v_{b0} \tau$ for the initial conditions \eqref{eq:IBC}. 
    }
    \label{fig:fig_profil_OTSI}
\end{figure}

Figure~\ref{fig:fig_profil_OTSI} illustrates the shape of the above solution as a function of $\theta$ for fixed $\Gamma_0 \tau = 5$. We observe a clear transition from a space-time behavior for $\theta \lesssim 1$ to a purely temporal growth for $\theta \gtrsim 1$. This can be understood as follows. When the saddle point and the pole are remote enough from each other such that $\sqrt{\tau} \vert \Delta (\theta) \vert \gg 1$, i.e., $\sqrt{\tau}\left[1-3(\frac{1-\theta}{2})^{2/3} \theta^{1/3} \right]^{1/2} \gg 1$, one has $\mathrm{erfc}\,\left[\sqrt{\tau} \vert \Delta (\theta) \vert e^{i\frac{\pi}{12}} \right] \simeq \frac{e^{-\tau \vert \Delta (\theta) \vert^2 e^{i\pi/6} - i \pi/12}}{\sqrt{\pi \tau} \vert \Delta (\theta)\vert}$, and so the $\mathrm{erfc}$ term in the r.h.s of Eqs.~\eqref{eq:np_final_1} and \eqref{eq:np_final_3} cancels out the second term between brackets. One then obtains the limiting expressions
\begin{align}
\delta n_p(\tau,\xi) &\simeq \frac{S}{2^{2/3}\sqrt{3\pi \Gamma_0 \tau}}\frac{(1-\theta)^{2/3}} {\theta^{1/6} (1-3\theta)} \exp\left[3 \Gamma_0 \tau \left( \frac{1-\theta}{2} \right)^{2/3} \theta^{1/3}e^{i\pi/6} - i\frac{\pi}{12}\right] \,, \quad 0 < \theta < 1/3 \,, \label{eq:np_spatiotemp}\\
    \delta n_p(\tau,\xi) &\simeq \frac{S}{3} \exp\left(\Gamma_0 \tau e^{i\pi/6}\right) \,, \quad \theta > 1/3 \,, \label{eq:np_temp}
\end{align}
when $\sqrt{\tau} \vert \Delta (\theta) \vert \gg 1$. Equation~(3) in the main text follows from further assuming $\theta \ll 1$ in Eq.~\eqref{eq:np_spatiotemp}.  
One can see that, as expected, the instability shares the same spatiotemporal evolution as the Green's function for $\xi \ll v_{b0}\tau$, while it grows in a purely temporal manner for $\xi \gtrsim v_{b0}\tau/3$. A similar mix of spatiotemporal and temporal growth regimes was found for the CFI under the same set of initial and boundary conditions \cite{Pathak_NJP_2015}.

\section{Evolution of the oblique instability in electron-beam-plasma simulations}

In the main text, we benchmark the spatiotemporal theory of the OTSI against
%To check the validity of the spatiotemporal theory of oblique modes, we performed
2D PIC simulations in which a semi-infinite, relativistic electron-positron pair beam propagates in a uniform plasma (see Fig.~2 of the main text). The reason behind using a pair beam rather than an electron beam is to suppress plasma wakefields and other sources of noise induced by a nonneutral charged particle beam, which hamper the observation of the OTSI evolution over a large dynamic range.

\begin{figure}[h!]
    \centering
    \includegraphics[width=0.99\textwidth]{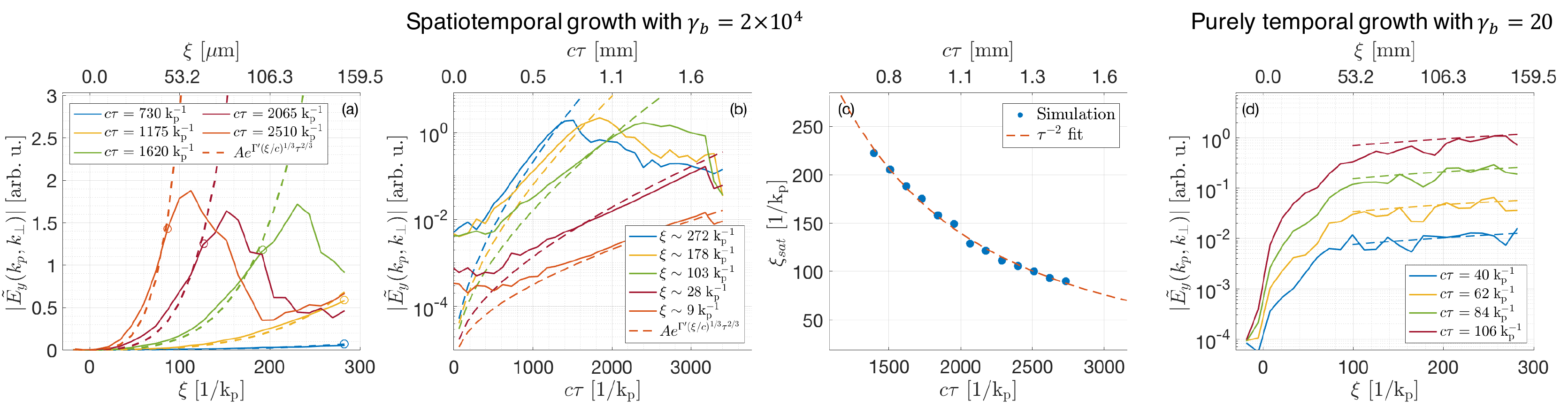}
    \caption{Same figures as Fig.~2 of the main text but with an electron beam instead of a neutral pair beam.}
    \label{fig:fig1_sup}
\end{figure}

Despite this complication, however, it remains possible to capture the spatiotemporal behavior of the instability in an electron beam-plasma simulation 
as is shown in Fig.~\ref{fig:fig1_sup}. Except for the pair beam being replaced with an electron beam of halved relative density ($n_b/n_p = 0.03$ vs. $0.06$ for the pair beam), the simulation parameters are identical to those of Fig.~2 in the main text. Although the initial plasma disturbances are stronger and vary appreciably with the longitudinal coordinate $\xi$, one still clearly observes the spatiotemporal character of the linear phase of the instability in the ultrarelativistic regime [$\gamma_b=2\times 10^4$, see Figs.~\ref{fig:fig1_sup}(a)-(c)], with reasonable agreement with the cold-fluid model. Similarly, the expected transition to a purely temporal growth at increasing distance from the beam edge can be observed at more moderate Lorentz factors, as illustrated by Fig.~\ref{fig:fig1_sup}(d) for $\gamma_b=20$.
Note that in the latter simulation, the dominant OTSI mode has a lower transverse wavenumber ($k_\perp \simeq 1.3 k_p$) than that ($k_\perp \simeq 3 k_p$) found in all the other simulations, a difference that is taken into account in computing the theoretical curves (dashed lines) of Fig.~\ref{fig:fig1_sup}(d).

\section{Competition between the spatiotemporal oblique instability and beam self-focusing}
\subsection{Comparison of characteristic time scales}
\begin{figure}[h!]
    \centering
    \includegraphics[width=0.43\textwidth]{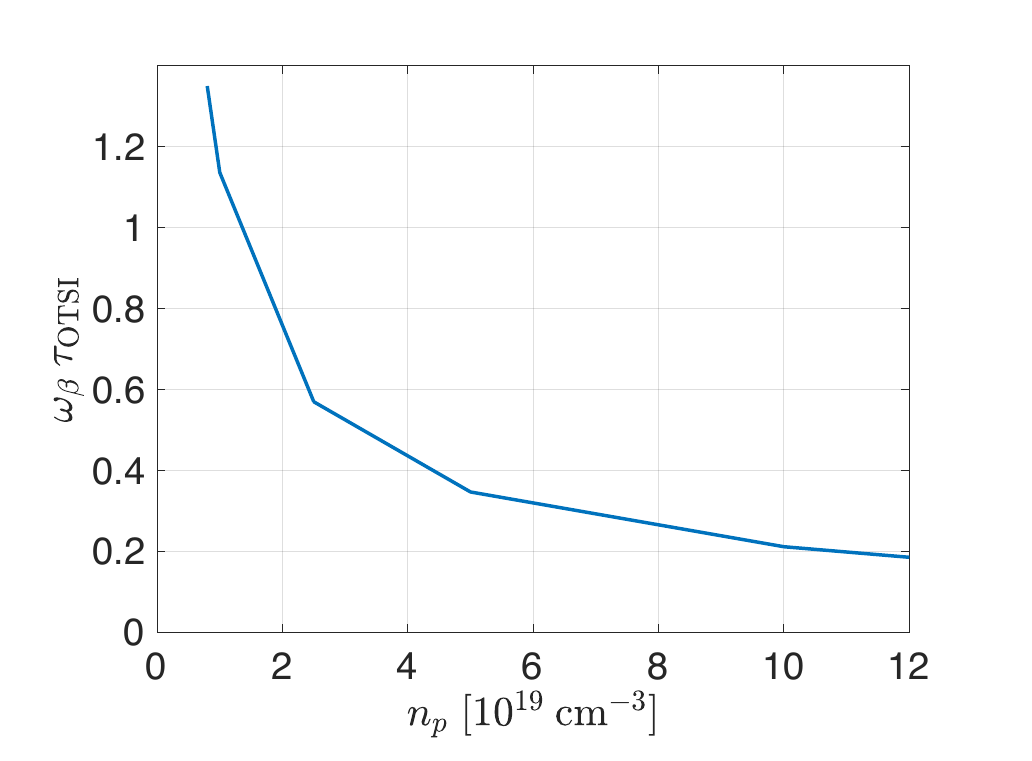}
    \caption{Ratio of the characteristic time scales of the OTSI ($\tau_{\rm OTSI}$) and beam self-focusing ($\omega_\beta^{-1}$) in the ultrarelativistic beam limit. The beam parameters are those associated with Fig.~3 of the main text. 
    }
    \label{fig:fig3_sup}
\end{figure}

In the main text, we show that the plasma wakefields driven by finite-size electron beams cause the resulting beam self-focusing to compete with the OTSI. The interplay of these two processes can be gauged by comparing their respective characteristic time scales. The time scale for beam self-focusing is estimated to be the inverse betatron frequency $\omega_\beta^{-1}=\sqrt{\gamma_b m_e/\partial_r W_\perp}$, where $W_\perp$ is the focusing force associated with the plasma wakefield in the linear regime. In the ultrarelativistic beam limit ($\gamma_b \gg 1$), $W_\perp$ is given by \cite{Keinigs_POF_1987}:
\begin{equation}
    W_\perp(\xi,r) = 4\pi k_p e \int_0^{\infty}dr'r'I_1(k_pr_<)K_1(k_pr_>) \int_{-\infty}^\xi d\xi' \partial_{r'} n_b(\xi',r')\sin(k_p(\xi-\xi')) \,,
    \label{eq:keinigs}
\end{equation}
where $I_1$ and $K_1$ are modified Bessel functions, and $r_<$ (resp. $r_>$) denotes the minimum (resp. maximum) of $r$ and $r'$. The radial derivative of $W_\perp$ can be numerically computed (at the beam center and $r=0$) for a Gaussian beam with RMS length $\sigma_x$ and width $\sigma_r$.

Since the electron beam being considered is short enough that $\Gamma_{\rm OTSI} \sigma_x/c \ll 1$ [$\Gamma_{\rm OTSI}$ is the maximum growth rate given by Eq.~(1) of the main text], the OTSI develops in a spatiotemporal manner. Consequently, its effective growth time, $\tau_{\rm OTSI}$, can be estimated from the leading exponential term of Eq.~(3) in the main text. Taking $\tau_{\rm OTSI}$ to be the time needed for a given beam slice, located at $\xi$, to experience a certain number $N_{\exp}$ of $e$-foldings, one obtains
\begin{equation}
    \tau_{\rm OTSI} = 2\left (\frac{N_\mathrm{exp}}{3\Gamma_\mathrm{OTSI}} \right)^{3/2} \left(\frac{v_b}{\xi}\right)^{1/2} \,,
    \label{eq:tau_OTSI}
\end{equation}
To get an approximate value for the Gaussian beam considered in Fig.~3 of the main text, we take $\xi = \sigma_x$, $n_b = \max n_b$, $k_\perp = 2k_p$ and $N_{\exp} = 5$ (consistent with the observed dynamic range of the growing field amplitude).

Figure~\ref{fig:fig3_sup} plots $\omega_\beta\, \tau_{\rm OTSI}$ as a function of the background plasma density for the beam parameters of Fig.~3 in the main text. One can see that $\omega_\beta \,\tau_{\rm OTSI} > 1$, meaning that beam self-focusing prevails over the OTSI, for $n_p \lesssim 10^{19}\,\rm cm^{-3}$, in good agreement with the simulation results (see main text). 

\subsection{Comparison of 2D and 3D simulations with a beam of finite length and width}

\begin{figure}[h!]
    \centering
    \includegraphics[width=0.7\textwidth]{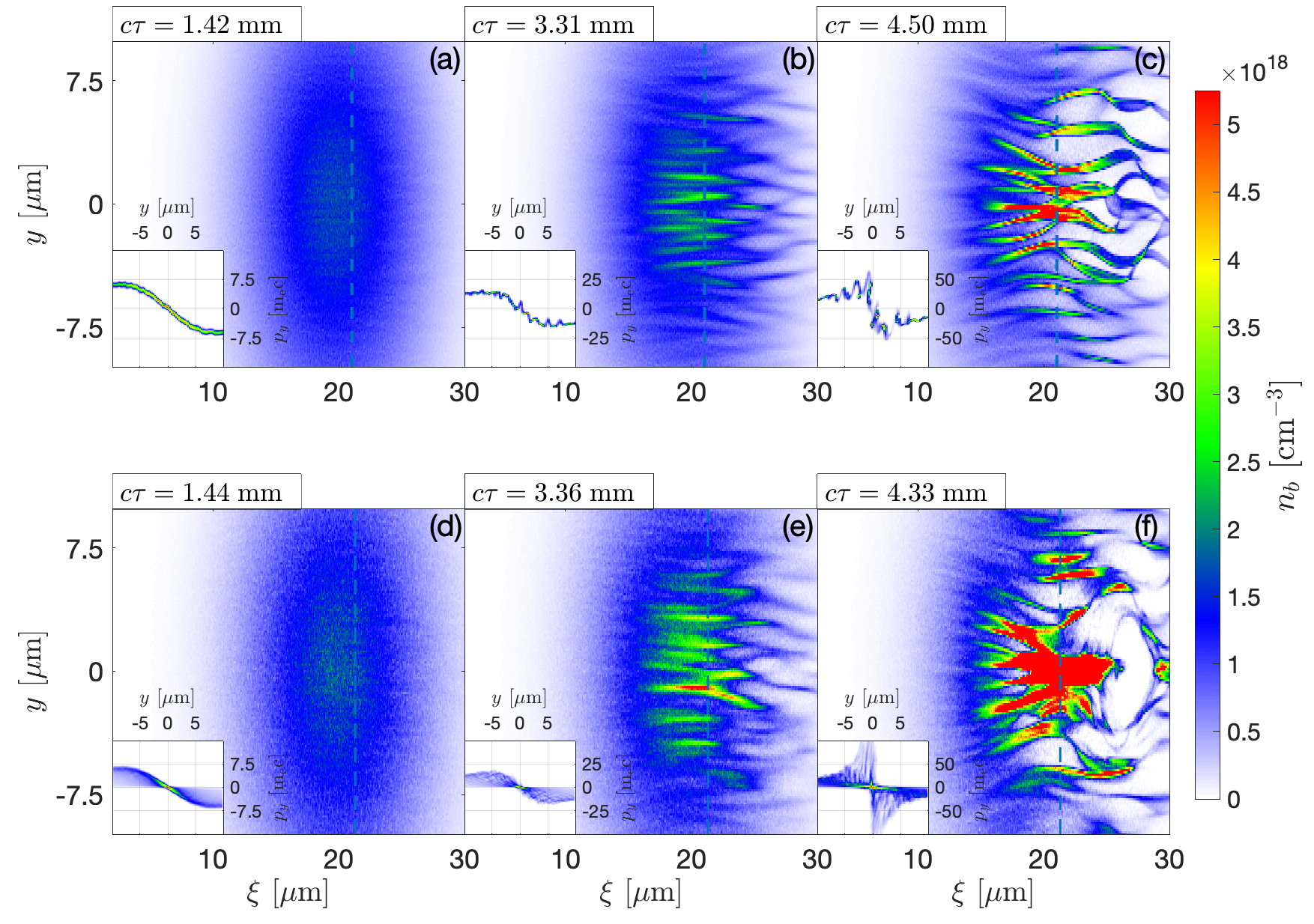}
    \caption{Simulated electron beam density maps at different propagation distances
in a uniform plasma of density $n_p=2.5\times 10^{19}\,\rm cm^{-3}$ in 2D [(a)-(c)] and 3D [(d)-(f)] geometry.
The insets show the transverse ($y,p_y$) beam phase space along
the dashed blue lines.
    }
    \label{fig:fig4_sup}
\end{figure}

The transverse wakefield given by Eq.~\eqref{eq:keinigs} applies to an axisymmetric beam in 3D geometry whereas the simulations shown in Fig.~3 of the main text were run in 2D. To appraise the impact of multidimensional effects on the interplay of the beam-plasma instability and beam self-focusing in a realistic setting, we performed a 3D simulation using the physical parameters of Figs.~3(g)-(i). The mesh size of this simulation was the same as in 2D but due to its higher computational cost, the number of macroparticles per cell was reduced to 4, 3 and 1 for, respectively, the beam electrons, plasma electrons and plasma ions. Figure~\ref{fig:fig4_sup} compares the beam density maps as extracted from the 2D [Figs.~\ref{fig:fig4_sup}(a)-(c)] and 3D [Figs.~\ref{fig:fig4_sup}(d)-(f)] simulations at different propagation distances. Panels (a)-(c) display the same plots as in Figs.~3(g)-(i) of the main text, while panels (d)-(f) show 2D slices (at $z=0$) from the 3D simulation. 
The insets depict the corresponding transverse ($y,p_y$) beam phase spaces along the slices indicated by the dashed blue lines. In 3D, these phase spaces are integrated over the $z$ direction, which explains why their spread is larger than in 2D.

We observe that the two simulations give overall comparable results. This is particularly true in the first 3.5 mm of propagation, at which point the instability saturates: the beam particles show the same amount of transverse phase-space rotation as a result of self-focusing [see insets of Figs.~\ref{fig:fig4_sup}(a)-(b) and (d)-(e)] and their density profile exhibits a similarly modulated pattern. We have also checked that the electromagnetic field energies grow similarly in both simulations. At later times, though, the instability enters its nonlinear stage, leading to significant differences between the 2D and 3D simulations. Notably, the 3D geometry allows more beamlets to merge into a single, larger-scale filament, leading to a faster transition to the blowout wakefield regime [compare Figs.~\ref{fig:fig4_sup}(c) and (f)]. A thorough analysis of these nonlinear phenomena lies outside the scope of this paper (focused on the linear stage of the beam-plasma interaction), but will be the subject of future work.

\providecommand{\noopsort}[1]{}\providecommand{\singleletter}[1]{#1}%